# Study of the nucleon radiative captures $^8$Li(n, γ)$^9$Li, $^9$Be(p, γ)$^{10}$B, $^{10}$Be(n, γ)$^{11}$Be, $^{10}$B(p, γ)$^{11}$C, and $^{16}$O(p, γ)$^{17}$F at thermal and astrophysical energies

Sergey Dubovichenko[*] and Albert Dzhazairov-Kakhramanov[†]

Fesenkov Astrophysical Institute "NCSRT" ASC MID Republic of Kazakhstan (RK), 050020, Observatory 23, Kamenskoe plato, Almaty, RK
[*]dubovichenko@mail.ru
[†]albert-j@yandex.ru

**Abstract:** We have studied the neutron-capture reactions $^8$Li($n$,γ)$^9$Li and its role in the primordial nucleosynthesis. The $n$+$^8$Li→$^9$Li+γ reaction has a significant astrophysical interest, because it includes into one of the variants of chain of primordial nucleosynthesis processes of the Universe and thermonuclear reactions in type II supernovae. Furthermore, we consider the $^9$Be(p,γ)$^{10}$B reaction in the astrophysical energy range in the modified potential cluster model with splitting of orbital states according to Young tableaux and, in some cases, with forbidden states. The reaction $^9$Be(p,γ)$^{10}$B plays an important role in primordial and stellar nucleosynthesis of light elements in the $p$ shell. Hydrogen burning in second-generation stars occurs via the proton–proton (pp) chain and CNO cycle, with the $^9$Be(p,γ)$^{10}$B reaction serving as an intermediate link between these cycles.

Furthermore, the possibility of describing available experimental data for the total reaction cross sections of neutron radiative capture on $^{10}$Be at thermal and astrophysical energies has been shown. This reaction is a part of one of the variants of the chain of primordial nucleosynthesis of the Universe due to which the elements with a mass of A > 11–12 may be formed. The results in the field of study of thermonuclear proton capture reaction on $^{10}$B at ultralow, i.e., astrophysical energies will be presented further. The possibility of description of the experimental data for the astrophysical $S$-factor of the proton radiative capture on $^{16}$O to the ground state of $^{17}$F was considered in the frame of the modified potential cluster model with forbidden states and classification of the states according to Young tableaux. It was shown that on the basis of the $E$1 transitions from the states of p$^{16}$O scattering to the ground state of $^{17}$F in the p$^{16}$O channel generally succeed to explain the value of measured cross sections at astrophysical energies.

## 1. Introduction

This review is the logical continuation of our works devoted to the radiative proton and neutron capture on light nuclei that were published in Int. J. Mod. Phys. E **21**, 1250039 (2012); **22**, 1350028 (2013); **22**, 1350075 (2013); **23**, 1430012 (2014).

### 1.1. *Astrophysical aspects of the review*

As we know, light radioactive nuclei play an important role in many astrophysical environments. In addition, such parameter as cross section of the capture reactions as a function of energy and reaction rates are very important for investigation of many astrophysical problems such as primordial nucleosynthesis of the Universe, main trends of stellar evolution, novae and super-novae explosions, X-ray bursts etc. The continued interest in the study of processes of radiative neutron capture on light nuclei at thermal (> 10 meV) and astrophysical (> 1 keV) energies is caused by several reasons. Firstly, this process plays a significant part in the study of many fundamental properties of nuclear reactions, and secondly, the data on the capture cross sections are widely used in a various applications of nuclear physics and nuclear astrophysics, for example, in the process of studying of the primordial nucleosynthesis reactions.

The n+$^8$Li→$^9$Li+γ reaction has a significant astrophysical interest, because it includes into one of the variants of chain of primordial nucleosynthesis processes of the Universe and thermonuclear reactions in type II supernovae.[1] In the process, primordial nucleosynthesis can occur after the nuclear formation with $A = 7$ the further way of formation of the elements with $A = 11$ and higher, for example, with help of the $^7$Li(n,γ)$^8$Li(α,n)$^{11}$B reaction or due to less likely branch $^7$Li(α,γ)$^{11}$B.[2,3] However, the neutron capture on $^8$Li, i.e., the n+$^8$Li→$^9$Li+γ reaction can reduce the amount of these nuclei in the given chain of fusion;[4] this reduction can reach the value of 50%. In other words, there are at least two variants of the primordial synthesis of elements with mass of $A = 11$–12 and greater. Notably, such a synthesis can take place not only due to the chain $^7$Li(n,γ)$^8$Li(α,n)$^{11}$B(n,γ)$^{12}$B(β$^+$)$^{12}$C, but in another way, for example, according to the reactions $^7$Li(n,γ)$^8$Li(n,γ)$^9$Li(α,n)$^{12}$B(β$^+$)$^{12}$C.[5] The reaction considered here $^8$Li(n,γ)$^9$Li is the basis of this chain.

It is possible to "produce" heavy isotopes using the *r*-process for supernovae stars of type II, immediately after the collapse stage. In particular, at the beginning of the expansion stage, the gap mass of elements at $A = 8$ can be eliminated by the reactions of the type α+α+α→$^{12}$C or α+α+n→$^9$Be.[6] However, overcoming such a mass gap is possible through the chain of reactions $^4$He(2n,γ)$^6$He(2n,γ)$^8$He(β$^-$)$^8$Li(n,γ)$^9$Li(β$^-$)$^9$Be on this stage,[7,8] where the process considered here is also presented. This chain would provide an alternative way toward heavier isotopes such as $^{36}$S, $^{40}$Ar, $^{46}$Ca, and $^{48}$Ca. However, it is very important to know to what extent this chain can compete with the process $^8$Li(β$^-$)$^8$Be(2α). An important answer to this question depends on knowing, as accurately as possible, the rate of the $^8$Li(n,γ)$^9$Li reaction and the neutron density.[1]

Inhomogeneous big bang models[9] predict a relatively higher prevalence of nuclides with $A > 8$ than the standard model. An unconvincing agreement[10] between primordial prevalence of the elements and those predicted with the standard model may be a starting point for the need for inhomogeneous models, in which $^7$Li(n,γ)$^8$Li(α,n)$^{11}$B are generally thought to be the major reaction chains forming heavier nuclei. However, it has been found that $^7$Li(n,γ)$^8$Li(n,γ)$^9$Li(α,n)$^{12}$B may be even more important,[11] because the $^8$Li(n,γ)$^9$Li reaction affects not only the reaction path to A > 8 isotopes but also the abundances of Li, Be, B, and C.

The problem in the investigation of the reaction $^8$Li(n,γ)$^9$Li is that the direct experimental measurement of the cross sections is impossible, because the half-life of $^8$Li is too small – 838 ms.[5] Microscopic calculations allowed us to obtain the rate of this reaction;[12,13] however such results differ by about an order of magnitude. For instance, in the comparative table of the results for the neutron capture reaction rate on $^8$Li from,[14] values from 790–5300 cm$^3$s$^{-1}$mole$^{-1}$ are given. However, this difference can be even greater, as shown further in Table 4. In addition, it is possible to produce a beam of $^9$Li nuclei, carrying out a measurement of the inverse reaction, $^9$Li+γ→$^8$Li+n, and using the principle of detailed balance to deduce the cross section for the neutron capture reaction. The photons for the inverse reaction are obtained by passing the $^9$Li through the virtual photon field near the nucleus of a high-$Z$ element such as Pb.[15]

In addition, the fact that these results can also depend on the presence of dark energy and its concentration,[16] on the rate of growth of baryonic matter perturbations[17] or on the rotation of the early Universe,[18] should not be excluded. However, perturbations in the primordial plasma not only stimulate the process of nucleosynthesis,[19] but also kill it, for example, through the growth of the perturbations of non-baryonic matter of the Universe[20] because of the oscillations of cosmic strings.[21]

As a result, recently, experimental efforts were carried out to study the reaction $^8$Li(n,γ)$^9$Li with the help of indirect approaches, for example, Coulomb dissociation[22,23] and (d,p) transfer reactions[5] (or (p,γ) in work[24]). Consequently, the total cross sections of the neutron capture on $^8$Li were obtained,[22] and then we use them further for comparison to our theoretical calculations.

Unfortunately, available theoretical results differ by several orders of magnitude (see Table 4) and it is difficult to draw any defined conclusions about the correctness or incorrectness of the homogeneous or the inhomogeneous models of the big bang. However, the great ambiguities in the study of this reaction by different methods make it interesting as a subject for study. Therefore we will consider the neutron capture on $^8$Li reaction in the frame of the modified potential cluster model (MPCM),[25,26] and will define how the criteria of this model can properly describe the total cross sections for the neutron capture on $^8$Li at thermal and astrophysical energies, namely from 25.3 meV to 1.0 MeV.

Some other calculation results of the total cross section of this reaction, based on different models and methods, which describe existing experimental data,[22] have been detailed elsewhere.[1,22] We could only find[27] experimental data of Ref. 22 and hardly any theoretical calculations of the cross sections; for example, Refs. 1, 12, 22 and some others, the results of which differ significantly from each other.

Furthermore, let us consider the p$^9$Be → $^{10}$Bγ reaction in the astrophysical energy range in the MPCM with splitting of orbital states according to Young tableaux and, in some cases, with forbidden states (FSs). We should note that we managed to find only one work devoted to a detailed experimental measurement of the cross sections and the astrophysical $S$-factor for this reaction[28] at low energies. We will use the results of this work furthermore for comparison with our model calculations.

The reaction $^9$Be(p,γ)$^{10}$B plays an important role in primordial and stellar nucleosynthesis of light elements in the $p$-shell.[29,30] Hydrogen burning in second-generation stars occurs via the proton–proton (pp) chain and CNO cycle, with the $^9$Be(p,γ)$^{10}$B reaction serving as an intermediate link between these cycles.

The knowledge of the p$^9$Be interaction potentials in continuous and discrete spectrum is required for the calculation of the astrophysical $S$-factor of the proton radiative capture on $^9$Be in the MPCM[31,32] which we usually used for the analysis of similar reactions.[33–35] Again we will consider that such potentials should correspond to the classification of cluster states according to orbital symmetries,[31,32,36] as it was assumed earlier for other light nuclear systems.

Continuing study of radiative capture processes,[37,38] let us consider in the frame of the MPCM the possibility of describing available experimental data for the total reaction cross sections and the reaction rate of the reaction n+$^{10}$Be→$^{11}$Be+γ at thermal and astrophysical energies. This reaction is a part of one of the variants of the chain of primordial nucleosynthesis of the Universe:[39]

$$…^{8}Li(n, γ)^{9}Li(β^{-})^{9}Be(n, γ)\underline{^{10}Be(n, γ)}^{11}Be(β^{-})^{11}B(n, γ)… \quad (1)$$

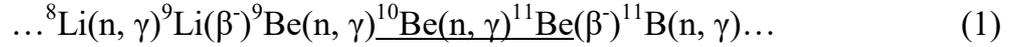

due to which the elements with a mass of A > 11–12 may be formed.[5]

In the next chapter the results in the field of study of thermonuclear proton capture reaction on $^{10}$B at ultralow, i.e., astrophysical energies will be presented. This reaction does not take part of thermonuclear cycles evidently, but up to now, evidently, was not considered on the basis of any model. In this case, as it was before, the MPCM is used as a nuclear model, which allows one to consider some thermonuclear processes, notably, radiative capture reactions of neutrons and protons on light nuclei, on the basis of unified views, criteria and methods.[40–43]

Continuing study of thermonuclear reactions in the frame of the MPCM with FSs[36] let us consider the $^{16}$O(p,γ)$^{17}$F process, which takes part of the CNO cycle[37] and has additional interest, since it is the reaction at the last nucleus of 1p-shell with the forming of $^{17}$F that get out its limit. As we usually assume,[37,38] the bound state (BS) of $^{17}$F is caused by the cluster channel of the initial particles, which take part in the reaction.

Many stars, including the Sun, will eventually pass through an evolutionary phase that is referred to as the asymptotic giant branch.[44] This phase involves a hydrogen and a helium shell that burn alternately surrounding an inactive stellar core. The $^{16}$O(p,γ)$^{17}$F reaction rate sensitively influences the $^{17}$O/$^{16}$O isotopic ratio predicted by models of massive (≥$4M_\odot$) AGB stars, where proton captures occur at the base of the convective envelope (hot bottom burning). A fine-tuning of the $^{16}$O(p,γ)$^{17}$F reaction rate may account for the measured anomalous $^{17}$O/$^{16}$O abundance ratio in small grains which are formed by the condensation of the material ejected from the surface of AGB stars via strong stellar winds.[45]

Earlier we have considered 27 thermonuclear processes, which can flow in the Sun, stars and in primordial nucleosynthesis of the Universe at different stages of its formation.[26,46–49] In the present studies of the radiative capture processes let us consider the n+$^{8}$Li→$^{9}$Li+γ, p$^{9}$Be → $^{10}$Bγ, n$^{10}$Be → $^{11}$Beγ, p$^{10}$B → $^{11}$Cγ, p$^{16}$O → $^{17}$Fγ reactions at thermal and astrophysical energies.

### 1.2. *Nuclear aspects of the review*

One extremely successful line of development of nuclear physics in the last 50-60 years has been the microscopic model known as the Resonating Group Method (RGM, see, for example,[50–54]). And the associated with it models, for example, Generator Coordinate Method (see, particularly,[54,55]) or algebraic version of RGM.[56,57] However, the rather difficult RGM calculations are not the only way in which to explain the available experimental facts. But, the possibilities offered by a simple two-body potential cluster model (PCM) have not been studied fully up to now, particularly if it uses the concept of FSs.[36] The potentials of this model for discrete spectrum are constructed in order to correctly reproduce the main characteristics of the BSs of light nuclei in cluster channels, and in the continuous spectrum they directly take into account the resonance behavior of the elastic scattering phase shifts of the interactive particles at low energies.[31,37] It is enough to use the simple PCM with FSs taking into account the described methods of construction of potentials and classification of the orbital states according to Young tableaux for consideration many problems of nuclear physics of low energy and nuclear astrophysics. Such a model can be called a modified PCM. In many cases, such an approach, as has been shown previously, allows one to obtain adequate results in the description of many experimental studies for the total cross sections of the thermonuclear reactions at low and astrophysical energies.[31,36,37]

Therefore, in continuing to study the processes of radiative capture,[31,37] we will consider the n+$^8$Li→$^9$Li+γ, p$^9$Be → $^{10}$Bγ, n$^{10}$Be → $^{11}$Beγ, p$^{10}$B → $^{11}$Cγ, p$^{16}$O → $^{17}$Fγ reactions within the framework of the MPCM at thermal and astrophysical energies. The resonance behavior of the elastic scattering phase shifts of the interacting particles at low energies will be taken into account. In addition, the classification of the orbital states of the clusters according to the Young tableaux allows one to clarify the number of FSs and allowed states (ASs), i.e., the number of nodes of the wave function (WF) of the relative motion of the cluster. The potentials of the n$^{10}$B interaction for scattering processes will be constructed based on the reproduction of the spectra of resonance states for the final nucleus in the n$^{10}$B channel. The n$^{10}$B potentials are constructed based on the description both of the binding energies of these particles in the final nucleus and of certain basic characteristics of these states; for example, the charge radius and the asymptotic constant (AC) for the BS or the ground state (GS) of $^{11}$B, formed as a result of the capture reaction in the cluster channel, which coincide with the initial particles.[31]

## 2. Model and calculation methods

The nuclear part of the intercluster interaction potential, which depends on set of quantum numbers *JLS*, for carrying out calculations of photonuclear processes in the considered cluster systems, has the form:

$$V(r,JLS) = -V_{JLS}\exp(-\gamma_{JLS}r^2), \qquad (2)$$

with point-like Coulomb term of the potential.

The potential is constructed completely unambiguously with the given number of BSs and with the analysis of the resonance scattering when in the considered partial wave at energies up to 1 MeV where there is a rather narrow resonance with a width of about 10–50 keV. Its depth is unambiguously fixed according to the resonance energy of the level at the given number of BS, and the width is absolutely determined by the width of such resonance. The error of its parameters does not usually exceed the error of the width determination at this level and equals 3–5%. Furthermore, it concerns the construction of the partial potential according to the phase shifts and determination of its parameters according to the resonance in the nuclear spectrum.

Consequently, all potentials do not have ambiguities and allow correct description of total cross sections of the radiative capture processes, without involvement of the additional quantity – spectroscopic factor $S_f$.[58] It is not required to introduce additional factor $S_f$ under consideration of capture reaction in the frame of PCM for potentials that are matched, in continuous spectrum, with characteristics of scattering processes that take into account resonance shape of phase shifts, and in the discrete spectrum, describing the basic characteristics of nucleus BS.

All effects that are present in the reaction, usually expressed in certain factors and coefficients, are taken into account at the construction of the interaction potentials. It could be possible, exactly because they are constructed and take into account FS structure. On the basis of description of observed, i.e., experimental characteristics of interacting clusters in the initial channel and formed, in the final state, a certain nucleus that has a cluster structure consisting of initial particles. In other words, the presence of $S_f$, is apparently taken into account in the BS WFs of clusters, determining the basis of such potentials due to solving the Schrödinger equation.[43]

The AC for any GS potential was calculated using the asymptotics of the WF having a form of exact Whittaker function[59]

$$\chi_L(r) = \sqrt{2k_0} C_w W_{-\eta L+1/2}(2k_0 r), \qquad (3)$$

where $\chi_L(r)$ is the numerical WF of the BS, obtained from the solution of the radial Schrödinger equation and normalized to unity, the value $W_{-\eta L+1/2}$ is the Whittaker function of the BS, determining the asymptotic behavior of the WF, which is the solution of the same equation without the nuclear potential. $k_0$ is the wave number, caused by the channel binding energy $E$: $k_0 = \sqrt{2\mu \frac{m_0}{\hbar^2} E}$; $\eta$ is the Coulomb parameter $\eta = \frac{\mu Z_1 Z_2 e^2}{\hbar^2 k}$, determined numerically $\eta = 3.44476 \cdot 10^{-2} \frac{\mu Z_1 Z_2}{k}$ and $L$ is the orbital angular momentum of the BS. Here $\mu$ is the reduced mass, and the constant $\hbar^2/m_0$ is assumed to be 41.4686 fm$^2$, where $m_0$ is the atomic mass unit (amu). The magnetic moment of neutron equals $\mu = -1.91304272\mu_0$, and $1.653560\mu_0$ for $^8$Li,[60] where $\mu_0$ is the nuclear magneton. Slightly transformed expressions[25] of Refs. 61,62 were used for total cross sections of the electromagnetic transitions.

Data on the spectroscopic factor $S$ of the GS and the asymptotic normalization coefficients $A_{NC}$ (ANC) are given, for example, in Ref. 5. Here we also use the relationship

$$A_{NC}^2 = S \times C^2, \qquad (4)$$

where $C$ is the asymptotic constant in fm$^{-1/2}$, which is related to the dimensionless AC $C_W$,[59] used by us in the following way: $C = \sqrt{2k_0} C_W$.

The total radiative capture cross sections $\sigma(NJ,J_f)$ for the $EJ$ and $MJ$ transitions in the case of the PCM are given, for example, in Ref. 58 or Refs. 31,37,63,64 are written as:

$$\sigma_c(NJ,J_f) = \frac{8\pi K e^2}{\hbar^2 q^3} \frac{\mu}{(2S_1+1)(2S_2+1)} \frac{J+1}{J[(2J+1)!!]^2}$$
$$\times A_J^2(NJ,K) \sum_{L_i,J_i} P_J^2(NJ,J_f,J_i) I_J^2(J_f,J_i) \qquad (5)$$

where σ – total radiative capture cross section; μ – reduced mass of initial channel particles; $q$ – wave number in initial channel; $S_1$, $S_2$ – spins of particles in initial channel; $K, J$ – wave number and momentum of γ-quantum in final channel; $N$ – is the $E$ or $M$ transitions of the $J$ multipole ordered from the initial $J_i$ to the final $J_f$ nucleus state.

The value $P_J$ for electric orbital $EJ(L)$ transitions has the form[31,37]

$$P_J^2(EJ,J_f,J_i) = \delta_{S_iS_f}\left[(2J+1)(2L_i+1)(2J_i+1)(2J_f+1)\right](L_i 0 J 0 | L_f 0)^2 \begin{Bmatrix} L_i & S & J_i \\ J_f & J & L_f \end{Bmatrix}^2,$$

$$A_J(EJ,K) = K^J \mu^J \left(\frac{Z_1}{m_1^J} + (-1)^J \frac{Z_2}{m_2^J}\right), \quad I_J(J_f,J_i) = \langle \chi_f | R^J | \chi_i \rangle. \qquad (6)$$

Here, $S_i$, $S_f$, $L_f$, $L_i$, $J_f$, and $J_i$ – total spins, angular and total moments in initial ($i$) and final ($f$) channels; $m_1$, $m_2$, $Z_1$, $Z_2$ – masses and charges of the particles in initial channel; $I_J$ –integral over WFs of initial $\chi_i$ and final $\chi_f$ states, as functions of cluster relative motion of n and $^{10}$B particles with intercluster distance $R$.

For consideration of the $M1(S)$ magnetic transition, caused by the spin part of magnetic operator,[65] it is possible to obtain an expression[31,37] using the following:[66]

$$P_1^2(M1,J_f,J_i) = \delta_{S_iS_f} \delta_{L_iL_f} \left[S(S+1)(2S+1)(2J_i+1)(2J_f+1)\right] \begin{Bmatrix} S & L & J_i \\ J_f & 1 & S \end{Bmatrix}^2,$$

$$A_1(M1,K) = \frac{e\hbar}{m_0 c} K\sqrt{3}\left[\mu_1 \frac{m_2}{m} - \mu_2 \frac{m_1}{m}\right], \quad I_J(J_f,J_i) = \langle \chi_f | R^{J-1} | \chi_i \rangle, \quad J=1. \qquad (7)$$

Here, $m$ is the mass of the nucleus, and $\mu_1$ and $\mu_2$ are the magnetic moments of the clusters, the values of which are taken from Refs. 67 and 68.

The construction methods used here for intercluster partial potentials at the given orbital moment $L$, are expanded in Refs. 37,31,69 and here we will not discuss them further. The next values of particle masses are used in the given calculations: $m_n$ = 1.00866491597 amu,[70] $m_p$ = 1.00727646577 amu,[71] $m(^8\text{Li})$ = 8.023540 amu,[72] $m(^9\text{Be})$ = 9.012182 amu,[73] $m(^{10}\text{Be})$ = 10.013533 amu,[74] $m(^{10}\text{B})$ = 10.012936 amu,[75] $m(^{16}\text{O})$ = 15.994915 amu,[76] and constant $\hbar^2/m_0$ is equal to 41.4686 MeV fm$^2$.

## 3. Neutron-capture reaction $^8$Li(n, γ)$^9$Li

### 3.1. Classification of n$^8$Li states according to Young tableaux

For $^8$Li, as well as for $^8$Be, accept the Young tableau {44}, because we have

{44} + {1} = {54} + {441} for the n$^8$Li system. It is clear that, in the state with $L = 0$, which we will need in the future, the FS with the tableau {54} presents. The AS corresponds to the configuration {441} at $L = 1$ – orbital angular momentum is determined by Eliot's rule.[77] The first of the obtained tableaux is forbidden, insofar as there are cannot be five nucleons in the s-shell, the second is allowed and is compatible with the orbital angular momentum $L = 1$, to which corresponds the GS of $^9$Li in the n$^8$Li channel.[78]

Thus, limited by the lower partial waves with the orbital angular momentum $L = 0,1$ it can be said that for the n$^8$Li system (for $^8$Li is known $J^\pi, T = 2^+, 1$ see Ref. 78) in the potentials of the P waves only the AS are present, while there are FS in the S waves. The state in $P_{3/2}$- wave corresponds to the GS of $^9$Li with $J^\pi, T = 3/2^-, 3/2$ and is located at the binding energy of the n$^8$Li system of -4.0639 MeV.[78] Some n$^8$Li scattering states and BSs can be mixed by spin with $S = 3/2$ ($2S + 1 = 4$) and $S = 5/2$ ($2S + 1 = 6$). However, here we will consider that the GS of $^9$Li in the n$^8$Li channel most probably is the $^4P_{3/2}$ level (in notations of $^{(2S+1)}L_J$), although in both spin states for $L = 1$ the total angular moment of $J = 3/2$ is possible.

In this case, we do not have the full tables of products of Young tableaux for a system with the number of particles greater than eight,[79] which were used by us earlier for such calculations.[25,47,80] Therefore, the obtained results should be considered only as a qualitative estimation of possible orbital symmetries in the GS of $^9$Li in the n$^8$Li channel. However, exactly on the basis of such a classification it was able to quite reasonably explain the available experimental data on radiative capture of neutrons and other particles for a wide range of nuclei of 1p-shell with masses from 6–16.[26,42,46–49,81,82] Therefore, here we will use the classification of cluster states by orbital symmetry, which leads us to a certain number of FSs and ASs in the partial intercluster potentials; this means that to a certain number of nodes of the relative motion WF of clusters: for the given case, neutron and $^8$Li. Even a qualitative estimation of orbital symmetries allows us to determine the presence of the FS in the S wave and the absence of the FS for the P states. Exactly the same structure of FSs and ASs in the different partial waves allows one to build the potentials of intercluster interaction that are necessary for calculations of the total cross section of the considered radiative capture reaction.

## *3.2. Structure of states in the n$^8$Li channel*

Now let us consider the structure of the excited and resonance states of $^9$Li in the n$^8$Li channel. We will take into account the first excited state (FES), which is bound in the n$^8$Li channel, and the first resonance state (FRS) of $^9$Li, because only for them the momenta and parities[78] are known. The FES of $^9$Li with $J^\pi = 1/2^-$ is located at the energy of 2.691(5) MeV relative to the GS or -1.3729(5) MeV relative to the threshold of the n$^8$Li channel. This state is the quartet $^4P_{1/2}$ level and its potential will have one bound AS.

The first resonance state is located at 4.296(15) MeV relative to the GS or at 0.2321(15) MeV relative to the threshold of the n$^8$Li channel. For this level the momentum $J^\pi = 5/2^-$ is given,[78] which allows one to take $L = 1$ for it, i.e., to consider it as a quartet $^4P_{5/2}$ resonance at 0.261(17) MeV in the laboratory system (l.s.). Its width of $\Gamma_{cm} = 100(30)$ keV is given in Ref. 78. It is possible to construct the quite unambiguous $^4P_{5/2}$ potential of the elastic scattering from these data. The ambiguities of its parameters will be caused only by the error of the width of this resonance, because it does not contain bound FSs or ASs.

On the basis of these data, one can consider that $E1$ capture is possible from the $^4S_{3/2}$ scattering wave to the $^4P_{3/2}$ GS of $^9$Li. As far as the GS with $^4P_{3/2}$ and the FES with $^4P_{1/2}$ we consider as the quartet states then the main contribution will give a transition of the form $^4S_{3/2} \rightarrow {}^4P_{3/2}$. For the radiative capture on the FES, the similar $E1$ transition $^4S_{3/2} \rightarrow {}^4P_{1/2}$ is possible. Since the results of the phase analysis of the elastic n$^8$Li scattering could not be found, several variants of the $^4S$ potential with the bound FS and different widths will be discussed. Such potentials should lead to $^4S$ phase shifts that are close to zero, as in the spectra of $^9$Li there are no $^4S$ resonances. Here, it should be noted that the FS will not necessarily be bound. This classification allows one to determine the presence of the FSs, but does not allow one to set whether it is bound.

Furthermore the GS potentials will be constructed for a correct description of the channel binding energy, the charge radius of $^9$Li and its AC in the n$^8$Li channel. Therefore, the known values of the ANC and the spectroscopic factor $S$, through which the AC is found, have quite a large error; the GS potentials will also have several variants with different parameters of width.

The radius of 2.327 ± 0.0298 fm was used in further calculations that are given in Ref. 83. We know the value of the radius of 2.2462 ± 0.0315 fm for $^9$Li from the database.[83] Moreover, for example, radii of these nuclei the next values are found: 2.299(32) fm for $^8$Li and 2.217(35) fm for $^9$Li.[84] In Ref. 85, for these radii were obtained the values 2.30(4) fm and 2.24(4) fm accordingly. All of these data within the error limits conform to each other. For the masses of the nucleus and neutron, the exact values were used: $^8$Li – 8.022487 amu[86] and $n$ – 1.00866491597 amu.[87] The charge radius of the neutron is assumed to be zero and its mass radius equals 0.8775(51) fm, which coincides with the known radius of a proton.[87]

## 3.3. The n$^8$Li interaction potentials

As in our previous works[25,47,80] we use the Gaussian potential (2). The GS of $^9$Li in the n$^8$Li channel is the $P_{3/2}$ level and this potential should describe the AC for this channel correctly. In order to extract the constant $C$ in the form (4) or $C_W$ (3) from the available experimental data, let us consider the information about the spectroscopic factors $S$ and the $A_{NC}$. For example, data that are given in Ref. 88 instead of our own results, and data for previous works, were relatively small for the considered system. To highlight these results from similar ones, i.e., with the close spectroscopic factor values, we present them in Table 1.

Furthermore, in two well-known papers the square of the ANC of the GS is determined – it equals Ref. 5, where $A^2_{NC} = 0.92(14)$ fm$^{-1}$ [$A_{NC} = 0.96(8)$ fm$^{-1/2}$] and in Ref. 24 is obtained as $A^2_{NC} = 1.33(33)$ fm$^{-1}$ [$A_{NC} = 1.15(14)$ fm$^{-1/2}$]. The average is equal to $A_{NC} = 1.06$ fm$^{-1/2}$ – this value agrees well with the results of Ref. 92, where for the GS the value of 1.08 fm$^{-1/2}$ with the values of spectroscopic factor from 0.58–1.02 (average 0.8) and the *ab initio* spectroscopic factor of 0.99 are given.[93] On the basis of the Equation (4) and the average $S$ for the AC of the GS from Table 1,[94] we find $C = 1.28$ fm$^{-1/2}$, and since $\sqrt{2k_0} = 0.915$, the dimensionless AC, defined as $C_W = C/\sqrt{2k_0}$, is equal to $C_W = 1.39(15)$.

Table 1. Data on spectroscopic factor $S$ for the GS of $^9$Li in the n$^8$Li from Refs. 88–91.

| Reaction from which spectroscopic factor $S$ is obtained | Spectroscopic factor values for the channel n+$^8$Li$_{g.s.}$ | Ref. |
| --- | --- | --- |
| $^2$H($^9$Li,$^3$H) | 0.65(15) | 88 |
| $^2$H($^9$Li,$^3$H) | 0.59(15) | 88 |
| $^2$H($^8$Li,$^1$H) | 0.90(13) | 89 |
| $^2$H($^8$Li,$^1$H) | 0.68(14) | 90 |
| $^9$Be($^8$Li,$^9$Li) | 0.62(7) | 91 |
| *Average value* | *0.69* | --- |
| *Data Limit* | *0.44–1.03* | --- |

However, if we use the upper value of the $A_{NC}$ equal to 1.29 (see Ref. 24) and the lower value of 0.88,[5] then at the average $S = 0.69$ for the dimensionless constant $C_W$ we will obtain a wider range of values of 1.43(27). One more interval of AC values can be estimated using the given range for the $A_{NC}$ and the interval for the spectroscopic factor value $S$, which is shown in Table 1. Then for the dimensionless $C_W$ we will obtain the wider range of values of 1.54(58).

Now we will note that for the ANC of the GS of $^9$Li in Ref. 95 at the spectroscopic factor of $S = 0.8$ the value $A_{NC} = 1.12$ fm$^{-1/2}$ was obtained, i.e., $C = 1.25$ fm$^{-1/2}$, which differs slightly from the average value of the ANC given above. Meanwhile, for the FES of $^9$Li in Ref. 95 the value of 0.4 fm$^{-1/2}$ at $S = 0.55$ was obtained, which leads us to $C = 0.54$ fm$^{-1/2}$ or to $C_W = 0.77$ at $\sqrt{2k_0} = 0.698$ in dimensionless form. In calculations Ref. 96 for the GS, the value of $A_{NC} = 1.140(13)$ fm$^{-1/2}$ was obtained, and for the FES the value of ANC is equal to $A_{NC} = 0.308(7)$ fm$^{-1/2}$, which, by and large, agree with all previous values.

Furthermore, we have suggested three variants of the GS potential, which allow one to obtain dimensionless asymptotic constant $C_W$, close to the value of 1.39. The parameters of these potentials $V_L$ and $\gamma_L$, and also basic characteristics of the nucleus obtained with them (the binding energy $E_{bnd}$, the asymptotic constant $C_W$, the mass radius $<R>_m$ and the charge radius $<R>_{ch}$) are listed in Table 2 of paper.[94]

All these potentials lead to the binding energy of -4.063900 MeV with an accuracy of the finite-difference method (FDM) used here for searching energy equal to $10^{-6}$ MeV. The determination of the calculated expressions for the radii is given, for example, in Refs. 25,26. The aforementioned AC errors are determined by averaging over the specified range of distances.

Table 2. Variants of the GS potential parameters and basic nuclear characteristics obtained with them.

| No. | $^{2s+1}L_J$ | $V_J$, MeV | $\gamma_J$, fm$^{-2}$ | Binding energy $E_{bnd}$, MeV | $C_W$ | $<R>_m$, fm | $<R>_{ch}$, fm |
|---|---|---|---|---|---|---|---|
| 1 | $^4P_{3/2}$ | 65.788593 | 0.18 | -4.063900 | 1.40(1) | 2.42 | 2.36 |
| 2 | $^4P_{3/2}$ | 71.714957 | 0.20 | -4.063900 | 1.31(1) | 2.41 | 2.35 |
| 3 | $^4P_{3/2}$ | 56.827345 | 0.15 | -4.063900 | 1.56(1) | 2.44 | 2.36 |
| 4 | $^4P_{1/2}$ | 56.727582 | 0.18 | -1.372900 | 0.80(1) | 2.53 | 2.37 |
| 5 | $^4P_{1/2}$ | 62.433328 | 0.20 | -1.372900 | 0.76(1) | 2.51 | 2.37 |

All potentials from Table 2 do not have the FSs in full accordance with the classification of states according to Young tableaux, given above. The GS potential $^4P_{3/2}$ (first variant from Table 2) was chosen exclusively for the correct description of the obtained AC value, equal to 1.39. The phase shift of the elastic scattering of this potential for the GS $^4P_{3/2}$ smoothly drop and at 1.0 MeV has the value in and around 174(1)°. However, as we have seen, the ANC and S have sizable errors; therefore another two variants of the GS potentials (2 and 3 variants from Table 2) that lead to similar results were considered.

The first of them, that is the variant of the potential from row 2 of Table 2, has the lower width, but leads to the same binding energy of -4.063900 MeV with an accuracy of 10$^{-6}$ MeV. The AC equals 1.31(1) on the interval 5–20 fm, the mass radius is 2.41 fm and the charge radius is 2.35 fm. The elastic scattering phase shift of this potential gradually decreases and at 1.0 MeV has the value of 174(1)°.

One more potential of the GS is wider than the previous one and has the parameters given in row 3 of Table 2. It allows one to obtain the same binding energy of -4.063900 MeV with an accuracy of 10$^{-6}$ MeV. The asymptotic constant $C_W$, mass radius $<R>_m$ and charge radius $<R>_{ch}$ are given in row 3 of Table 2. The elastic scattering phase shift gradually decreases and at 1.0 MeV has the value of 173(1)°. All these three potentials, in general, reflect the first interval errors of determination the dimensionless $C_W = 1.39(15)$, and the potential from row 3 of Table 2 agrees also with the latest estimation of the AC value, $C_W = 1.54(58)$.

The potential of the first excited $^4P_{1/2}$ state is obtained by the simple decreasing of the depth of the GS potential from row 1 of Table 2 and has the parameters given in row 4 of Table 2. The basic characteristics obtained with these parameters of the FES potential are listed in row 4 of Table 2. The value of the AC is not significantly greater than the value $C_W = 0.77$, obtained in Ref. 95. The elastic scattering phase shift gradually decreases and at 1.0 MeV has a value of 168(1)°.

There is another variant of the FES potential, obtained on the basis of the GS potential from row 2 of Table 2. The basic characteristics obtained with these parameters of the FES potential are listed in row 5 of Table 2. For this potential the AC value practically coincides with the results of Ref. 95. The elastic scattering phase shift gradually decreases and at 1.0 MeV has a value of 169(1)°.

Now let us consider the potential construction criteria for the $^4S$ scattering wave. First of all, as indicated above, such a potential must have the FS, which, however, should not necessarily be bound. In addition, since there are no results of the phase shift analysis of the n$^8$Li elastic scattering, and in the spectra of $^9$Li at the energy below 1.0 MeV there are no resonances of the positive parity, we assume that the $^4S_{3/2}$ potential should lead to almost zero phase shifts in this energy range.

The potential of the nonzero depth with the bound FS should be relatively narrow, to obtain smoothly varying scattering phase shifts close to zero. The parameters of such a potential are given in the first two rows of Table 3.[94]

Table 3. Variants of the potential parameters for unbound states of nucleus and certain characteristics obtained with them.

| No. | $^{2s+1}L_J$ | $V_J$, MeV | $\gamma_J$, fm$^{-2}$ | Resonance energy $E_{res}$ (l.s.), MeV | $\Gamma_{cm}$, keV |
|---|---|---|---|---|---|
| 1 | $^4S_{3/2}$ | 327.0 | 1.0 | --- | --- |
| 2 | $^4S_{3/2}$ | 653.5 | 2.0 | --- | --- |
| 3 | $^4P_{5/2}$ | 54.446 | 0.20 | 0.261(1) | 106(1) |

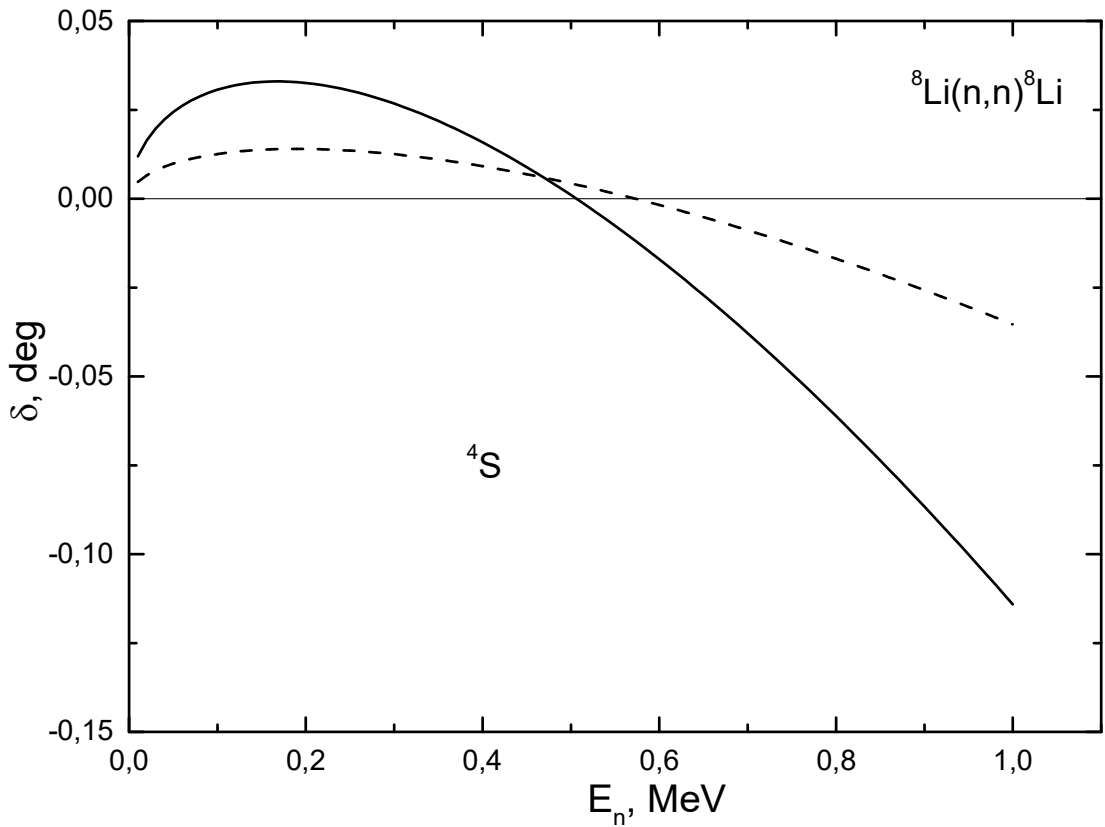

Fig. 1. The n$^8$Li elastic scattering phase shift in the $^4S$ wave at low energies. Lines are explained in the text.

The potential of row 1 in Table 3 leads to the scattering phase shifts that are shown in Fig. 1 by the solid line, which at energies up to 1.0 MeV is located in the range from +0.03° to –0.12°. The second variant of the scattering potential has the parameters given in row 2 in Table 3 and its phase shift is represented in Fig. 1 by the dashed line, having a value less than ±0.05° at the considered energy range. In addition to these two variants of the $^4S$ scattering potentials, the variance of such an interaction with zero depth will be considered, which leads to a zero phase shift and has no FS.

The potential without the bound FS for the $^4P_{5/2}$ scattering resonance has the parameters given in row 3 of Table 3 and leads to the resonance energy of 0.261(1) MeV (l.s.) at the width $\Gamma_{cm} = 106(1)$ keV with a scattering phase shift of 90.0°(1), which is in good agreement with Ref. 78.

### 3.4. Total cross sections and reaction rates of the neutron radiative capture on $^8Li$

As already mentioned, we will assume that the radiative capture for the $E1$ process comes from the $^4S_{3/2}$ scattering wave to the $^4P_{3/2}$ GS of $^9Li$ in the n$^8$Li channel.[25,26] The calculations of the total cross sections that we carried out for the potential of the GS from row 1 of Table 2 lead to the results shown in Fig. 2 by the solid line. The results for the potential of the GS from row 2 of Table 2 are shown by the dashed line and for the potential from row 3 of Table 2 by the dotted line. In all of these calculations, for the potential of the elastic scattering, the potential parameters from row 1 of Table 3 were used.

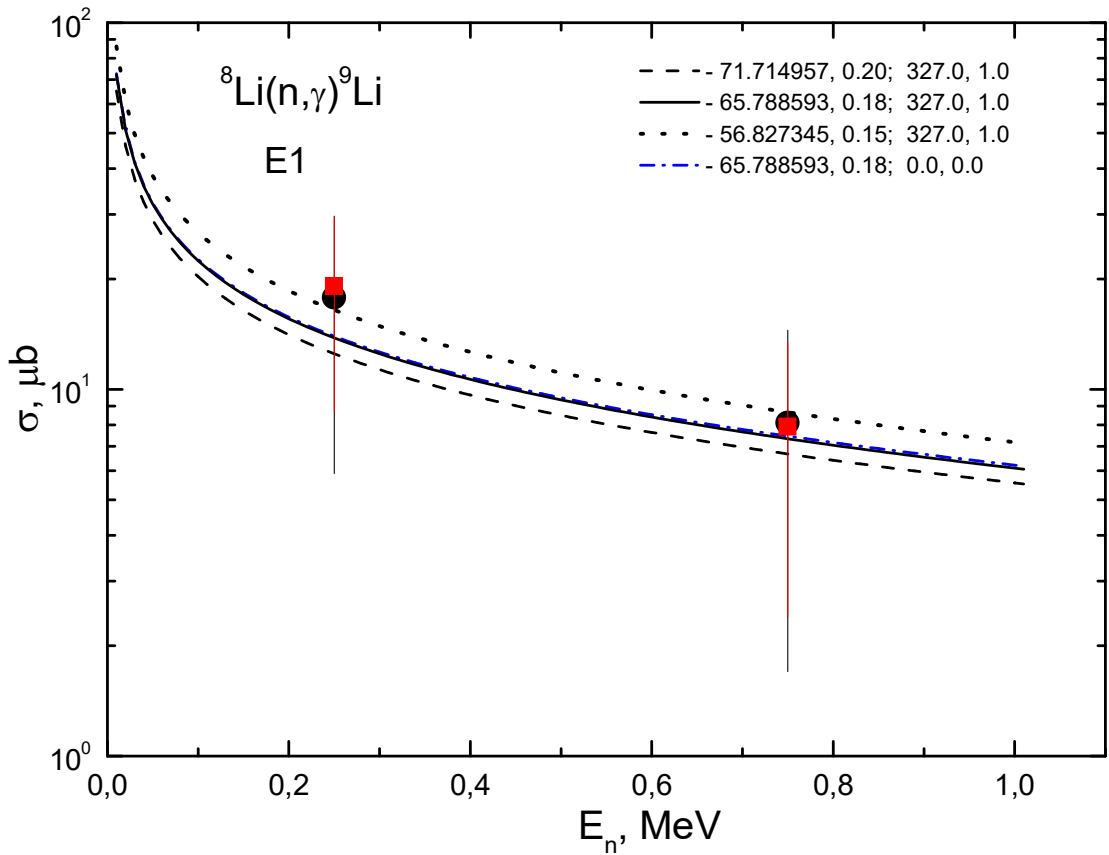

Fig. 2. The total cross sections of the $^8$Li(n,γ)$^9$Li radiative capture to the GS of $^9$Li. Black points – experimental data from Ref. 22 on Pb, red squares – data from Ref. 22 on U. Lines are explained in the text.

As we can see from these results, the shape of the cross sections weakly depends on the potential of the GS and all the proposed variants reproduce the available experimental data of Ref. 22 acceptably. For comparison, the results for the potential of the $^4S$ scattering wave of zero depth are given, i.e., without the FS, and for the potential of the GS from row 1 of Table 2 – blue dot-dashed line in Fig. 2. They do not significantly differ from the results of calculation of the cross sections for the scattering potential from row 1 of Table 3 with the same interaction of the GS from row 1 of Table 2, which are shown by the solid line.

Now we present the results of calculations of the total cross sections of the radiative capture for the scattering potential from row 2 of Table 3. All of them are shown in Fig. 3 – all lines have a marked similarly in Fig. 2. It can be seen from these results that the total capture cross section not only weakly depends on the potential of the GS, but practically does not depend on the shape and depth of the scattering potential, and such a potential does not necessarily have the bound FS – the equality of the phase shifts to zero is only important.

The results of calculation of the cross sections of Ref. 1 are shown in Fig. 3 by the green dot-dashed line, which are based on the Coulomb dissociation of $^9$Li. Note that results in Ref. 22 average over the energy ranges 0–0.5 MeV and 0.5–1.0 MeV. In addition, in Ref. 22, only the upper limit of the capture cross sections is given, therefore the results of Ref. 1 describe such experimental data quite accurately. Furthermore, the results of Ref. 13, based on a microscopic cluster model, are shown by the blue dot-dot-dashed line in Fig. 3. These results do not differ significantly from our own,[94] shown in Figs. 2 and 3 for the potential of the GS from row 2 of Table 2 by the dashed line.

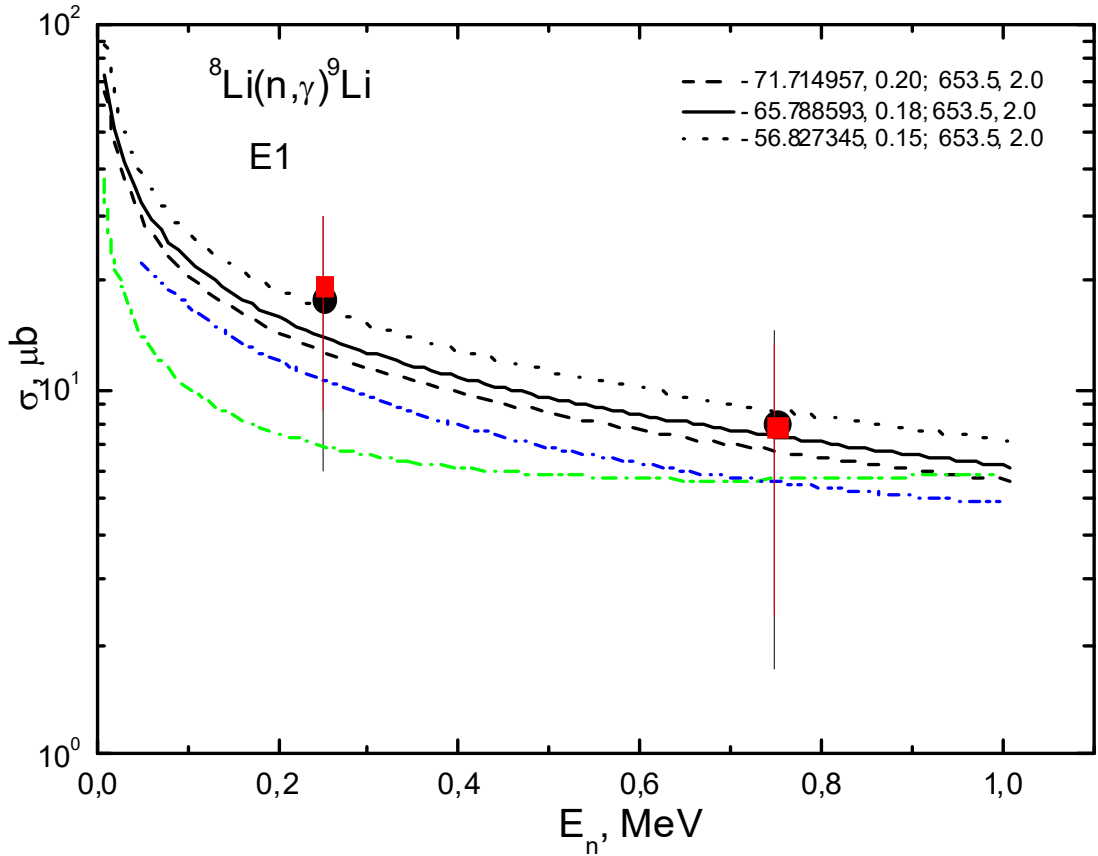

Fig. 3. The total cross sections of the $^8$Li(n,γ)$^9$Li radiative capture to the GS of $^9$Li. Black points – experimental data from Ref. 22 on Pb, red squares – data from Ref. 22 on U. Lines are explained in the text.

Our calculations[94] of the $E1$ cross section for the GS potential from row 2 of Table 2 and the scattering potential from row 2 of Table 3 at thermal energy of 25.3 meV lead to the value of 41.3 mb. The results of this calculation, as a function of energy in the region of 25 meV – 1.0 MeV, are shown in Fig. 4 by the solid line. For the same GS potential and the scattering potential from row 1 of Table 3 the thermal cross section has the value of 41.2 mb and for the scattering potential of the zero depth without FS, a value of 41.4 mb was obtained. In Ref. 13, for the total cross section value at thermal energy, 37.9 mb is given.

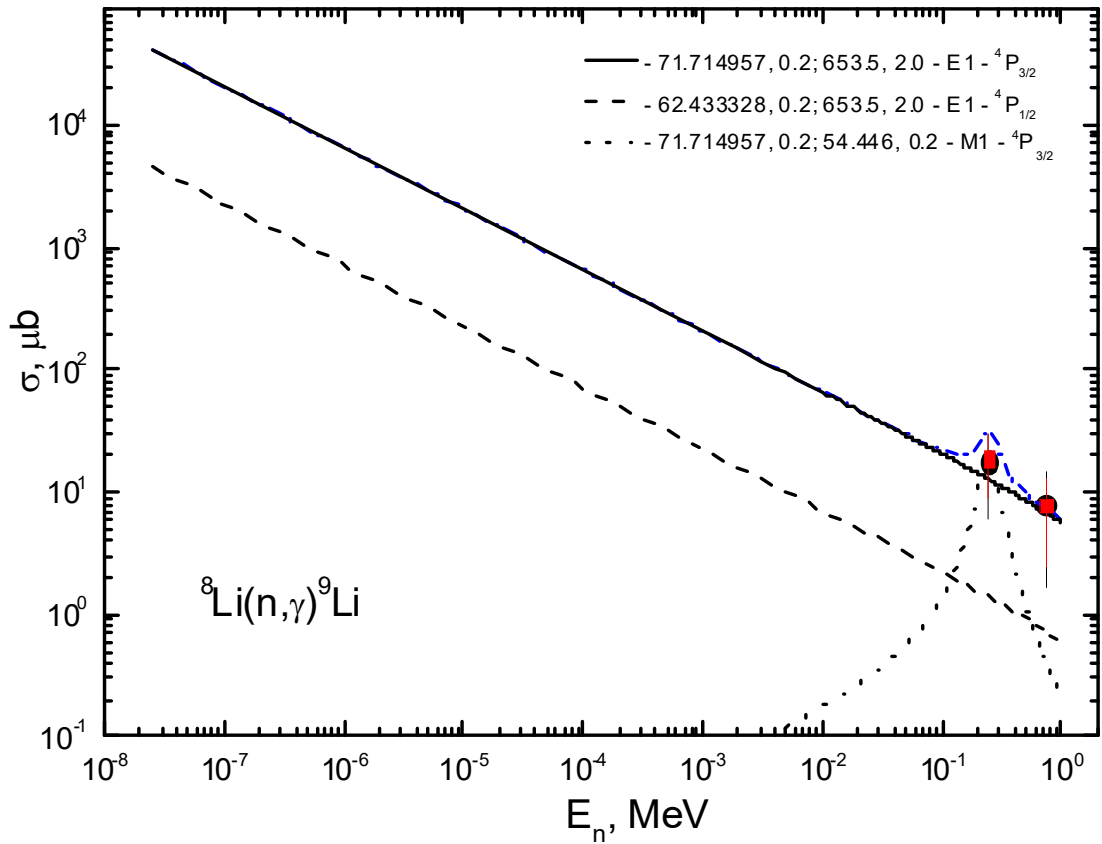

Fig. 4. The total cross sections of the $^8$Li(n,γ)$^9$Li radiative capture to the GS of $^9$Li. Black points – experimental data from Ref. 22 on Pb, red squares – data from Ref. 22 on U. Lines are explained in the text.

The $E$1 transition cross section to the FES with the potential from row 5 of Table 2 and the scattering potential from row 2 of Table 3 is shown in Fig. 4 by the dashed line, having a much smaller value of 4.5 mb at thermal energy of 25.3 meV. A rather greater value of 4.9 mb was obtained for the FES potential from row 4 of Table 2 and the scattering potential from row 2 of Table 3. This cross section has values of almost an order of magnitude smaller than that for the transition to the GS at the range of 0.1– 1.0 MeV.

The cross section of the $M$1 transition $^4P_{5/2} \to {}^4P_{3/2}$ from the resonance scattering wave with the potential from row 3 of Table 3 to the GS with the potential from row 2 of Table 2 is shown by the dotted line in Fig. 4. The blue dot-dashed line with the resonance in the area of 0.26 MeV is the summed total cross section of the $E$1 and $M$1 transitions to the GS. It can be seen from these results that the consideration of the $M$1 transition leads to a small resonance in the cross sections, which at such large measurement errors does not affect their magnitude significantly. Additional consideration of the $E$1 transition to the $^4P_{1/2}$ FES increases the summed total cross sections approximately by 10%, leading to thermal cross sections of about 46 mb.

Furthermore, in Fig. 5, the rate $N_A\langle\sigma v\rangle$ of the neutron capture on $^8$Li is shown by the solid red line, which corresponds to the solid line in Fig. 2 and is presented in the form[58]

$$N_A\langle\sigma v\rangle = 3.7313 \cdot 10^4 \mu^{-1/2} T_9^{-3/2} \int_0^\infty \sigma(E) E \exp(-11.605 E/T_9) dE, \qquad (8)$$

where $E$ is in MeV, the cross section $\sigma(E)$ is measured in µb, µ is the reduced mass in amu, $T_9$ is the temperature in $10^9$ K. Integration of cross sections was carried out in the range 0.1 keV – 2 MeV, and expansion of this interval led to a change of the reaction rate of about 1%.

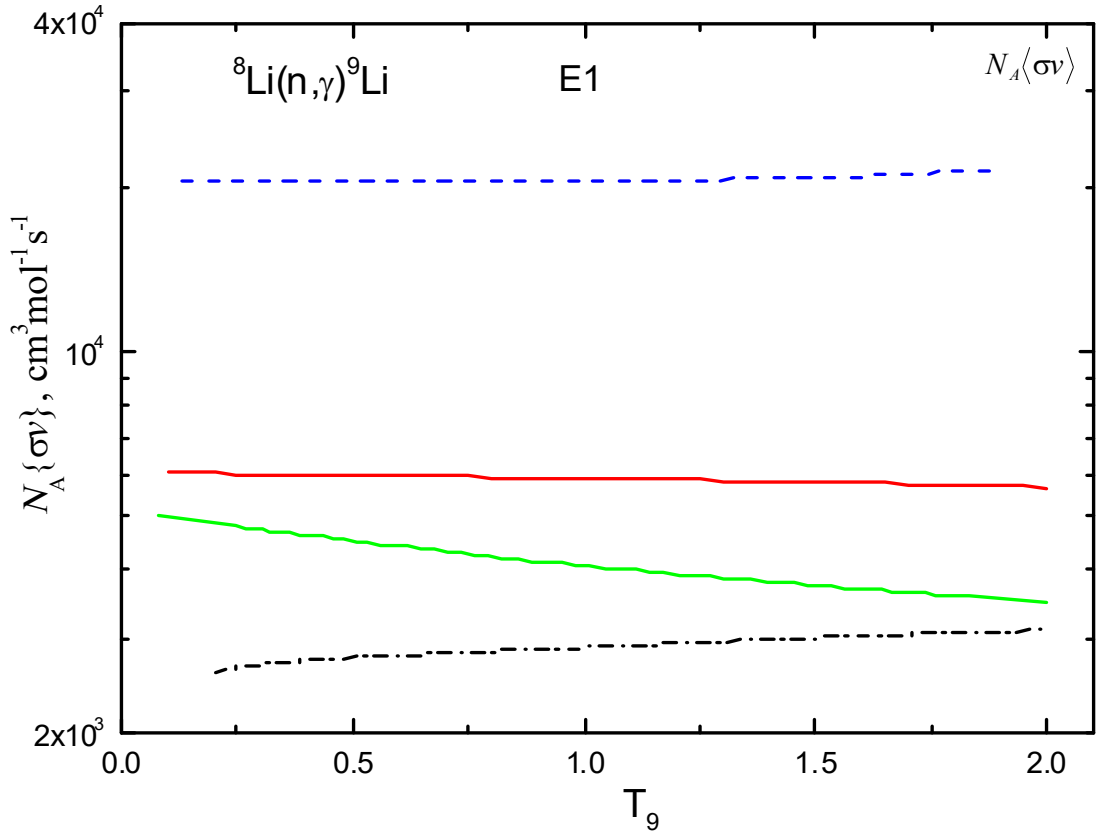

Fig. 5. Reaction rates of the neutron capture on $^8$Li. Lines are explained in the text.

The reaction rate for the results of Ref. 1 is shown by the black dot-dashed line in Fig. 5; the cross section for this work is shown in Fig. 3 by the green dot-dashed line. It is easy to see, from Figs. 3 and 5, the difference of the total cross sections and reaction rates, obtained in different works. In addition, in Fig. 5 the blue dashed line shows the results of reaction rate for the direct capture to the GS from Ref. 97, which is a few times greater than that obtained in this work. In Fig. 5 the green solid line shows the results of the direct capture to the GS from Ref. 90, which better coincides with our calculations.

Furthermore, the comparison of reaction rates for the $^8$Li$(n,\gamma)^9$Li reaction at $T_9 = 1$ reported by various workers is listed in Table 4.[94]

Table 4. Comparison of reaction rates for the direct capture of the $^8$Li$(n,\gamma)^9$Li reaction at $T_9 = 1$.

| Reference | Year of publication | Reaction rate (cm$^3$ mole$^{-1}$ s$^{-1}$) |
|-----------|---------------------|---------------------------------------------|
| 4         | 1988                | 43000                                       |
| 97        | 1991                | 21000                                       |
| 22        | 1998                | <7200                                       |
| *94*      | *2016*              | *5900*                                      |
| 13        | 1993                | 5300                                        |
| 11        | 1994                | 4500                                        |
| 90        | 2005                | 4000                                        |
| 1         | 2008                | 2900                                        |
| 12        | 1999                | 2200                                        |
| 23        | 2003                | <790                                        |

As can be seen Table 4, the rate of the $^8$Li$(n,\gamma)^9$Li reaction, obtained by us, differs from the results of other Refs. 11,13,90 by not more than 30%. In addition, our results are in a good agreement with the experimental data,[22] which at $T_9 = 1$ gives <7200. However, it must be noted that our value[94] of the reaction rate, as well as Ref. 22, differs considerably from the conclusions of Ref. 97, where calculations where performed in the *spd*-model, and Ref. 4, the results of which were obtained in the systematics of similar nuclei.

Since previously we had obtained all potentials of the $^4P$ waves, which can be used for calculation of the characteristics of the BSs and for calculation the phase shifts of the elastic scattering, they can be used for consideration of the $E2$ transitions from the scattering states to the GS:

$$\text{Process No.1. } {}^4P_{1/2} \to {}^4P_{3/2}$$
$$\text{Process No.2. } {}^4P_{3/2} \to {}^4P_{3/2}$$
$$\text{Process No.3. } {}^4P_{5/2} \to {}^4P_{3/2}$$

Also for the $E2$ capture to FES

$$\text{Process No.4. } {}^4P_{3/2} \to {}^4P_{1/2}$$
$$\text{Process No.5. } {}^4P_{5/2} \to {}^4P_{1/2}$$

The calculation results of the total cross sections for the $E2$ transitions from all $^4P$ scattering waves to the GS are shown in Fig. 6 by the red dot-dashed line. Results for process No. 1 with the scattering potential in the $^4P_{1/2}$ wave from row 5 of Table 2 and the GS potential from row 2 of Table 2 are presented by the dotted line. Process No. 2 with the GS potentials from row 2 of Table 2 in the continuous and discrete spectrum is depicted as the dashed line. Process No. 3 with the $^4P_{5/2}$ scattering potential from row 3 of Table 3 and the GS potential from row 2 of Table 2 is shown by the black solid line.

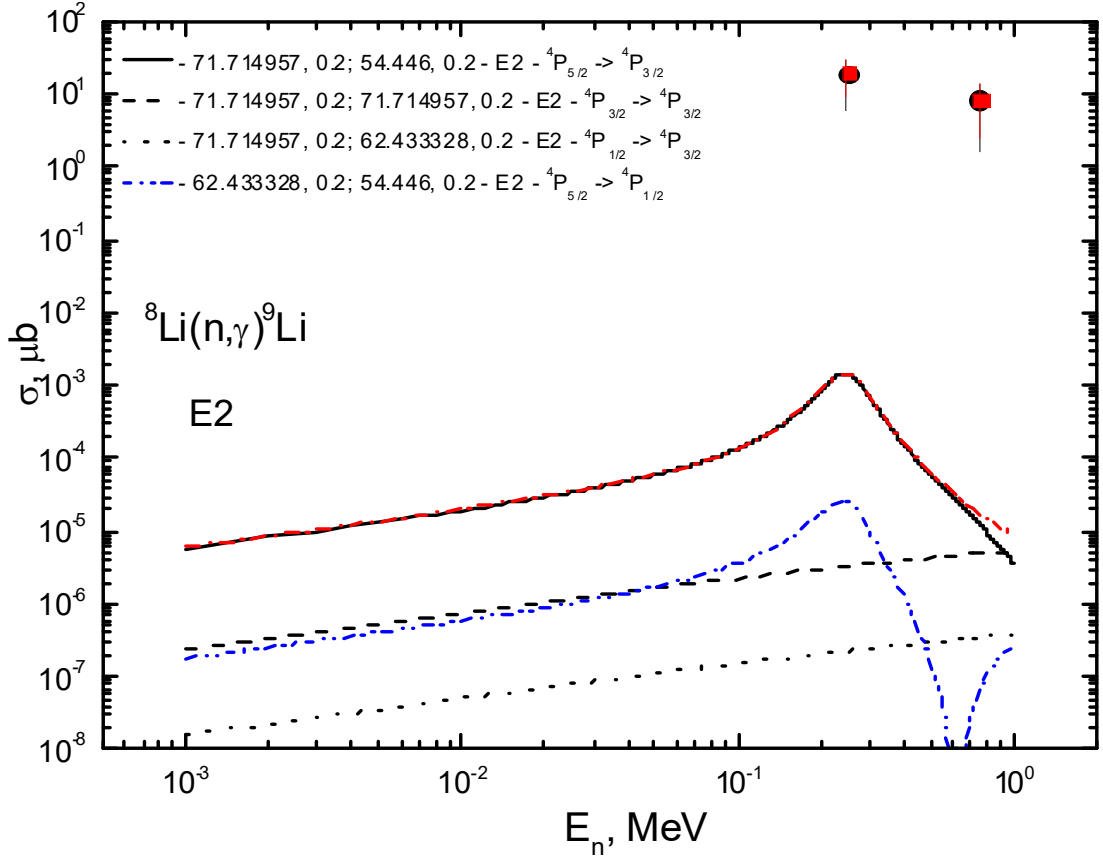

Fig. 6. The total cross sections of the $^8$Li(n,γ)$^9$Li radiative capture for the $E2$ transitions from all $^4P$ scattering waves to the GS of $^9$Li. Black points – experimental data from Ref. 22 on Pb, red squares – data from Ref. 22 on U. Lines are explained in the text.

The total cross sections for transitions to the FES have an even smaller value for resonance process No. 5 with the scattering potential from row 3 of Table 3, and the FES potential from row 5 of Table 2 is shown in Fig. 6 by the blue dot-dot-dashed line. It can be seen from these results that in the used MPCM with the potentials proposed above, the considered $E2$ transitions, even from the resonance $^4P_{5/2}$ wave, do not provide any contribution to the total cross sections of the neutron radiative capture on $^8$Li.

Since at the energies from 25.3 meV and up to ~100 keV, the calculated cross section is a straight line (i.e. the solid line in Fig. 4), it can be approximated by a simple function of energy of the form:[25,26]

$$\sigma_{ap}\,(\mu b) = \frac{A}{\sqrt{E_n\,(\text{keV})}}. \qquad (9)$$

The constant value of $A = 207.8550$ μb keV$^{1/2}$ was determined by one point in the calculated cross sections at the minimum energy of 25.3 meV. The modulus of relative deviation

$$M(E) = \left| [\sigma_{ap}(E) - \sigma_{theor}(E)] / \sigma_{theor}(E) \right| \qquad (10)$$

between the theoretically calculated cross section ($\sigma_{theor}$) and its approximation ($\sigma_{ap}$)

given by Equation (9) in the energy range up to 10 keV has a value of less than 0.2%, while for increasing energy up to 100 keV, it increases to 1.5–2.0%.

It is realistic to assume that this form of dependence of the total cross section from the energy will survive at lower energies. Therefore, it is possible to perform an estimation of the cross section, for example, at the energy of 1 μeV (1 μeV = $10^{-6}$ eV), which gives the value of the order of 6.6 b on the basis of the expression given above for approximation of the cross section in Equation (9).

If one uses the zero potential of scattering for calculation of the cross section, then the coefficient in Equation (9) for the approximation of the calculation results of the total cross section is equal to $A = 208.1156$ μb keV$^{1/2}$. Estimation of the accuracy of approximation of the calculated capture cross section by Equation (9) in this case is on the same level.

## 4. Proton-capture reaction $^9$Be(p, γ)$^{10}$B

### 4.1. *Structure of states in the p$^9$Be system*

First, we will define the orbital Young tableaux of $^9$Be, for example, in the p$^8$Li or in the n$^8$Be channels. If we assume that it is possible to use tableaux {44} + {1} in the 8 + 1 system, then two possible symmetries: {54} + {441} are obtained for this system. The first of them is forbidden because it contains five cells in one matrix row.[77] We note, promptly, that the given classification of orbital states according to Young tableaux has qualitative character, because there is no table of Young tableaux products for the system with $A = 9,10$ particles, though they exist for all systems with $A < 9$ (see Ref. 79) and they are used for the analysis of the number of ASs and FSs in WFs of different cluster systems.[64]

Furthermore, if the tableau {54} is used for $^9$Be then, possible, Young tableaux of the p$^9$Be system turn out to be forbidden, because of the rule that it can not be more than four cells in one row.[77,98] They are corresponded to FSs with configurations {64}, {55} and relative motion moments $L = 0$ and 1, which is determined by the Elliot rule.[77] Another forbidden tableau {541} is present in this product as in the examined case too and correspond to $L = 1$.

When the tableau {441} is accepted for $^9$Be, the p$^9$Be system contains FSs with the {541} tableau in the *P*-wave and, evidently, with the {442} in the *S*-wave and AS configuration {4411} with $L = 1,3$. Thus, the p$^9$Be potentials in different partial waves should have forbidden bound {442} state in the *S*-wave and forbidden and allowed bound levels in the *P*-wave with {541} and {4411} Young tableaux, respectively.

We can examine the case when for $^9$Be both possible orbital Young tableaux {54} and {441} are used. The same approach was quite successfully used earlier for consideration of the p$^6$Li (Ref. 99) and the p$^7$Li (Ref. 100) systems. Then the level classification will be slightly different – the number of FSs will increase and an additional forbidden bound level will appear in every partial wave with $L = 0$ and 1. Such a more complete tableau of states, which we will use later, equals the sum of the first and the second cases considered above and there are two FSs in the *S* and *P* waves with ASs in the *P* wave. One of them – the $^5P_3$ state – can correspond to the GS of $^{10}$Be in the p$^9$Be channel.

## 4.2. *Potential description of scattering phase shifts*

The considered p$^9$Be channel in $^{10}$Be has the isospin projection equal to $T_z = 0$, which is possible with two values of the total isospin $T = 1$ and $0$,[78] therefore the p$^9$Be system, as well as p$^3$H,[33] turns out to be isospin-mixed. In this case the p$^9$Be and n$^9$Be cluster channels with $T_z = \pm1$ and $T = 1$ are isospin-pure in a complete analogy with the p$^3$He and n$^3$H systems.[33] The phase shifts of the elastic p$^9$Be scattering are represented as a half-sum of pure in isospin phase shifts[32,36] because this system is isospin-mixed, as it was given above in the expression $\delta^{\{4\}+\{31\}} = 1/2\delta^{\{31\}} + 1/2\delta^{\{4\}}$.

The isospin-mixed phase shifts with $T = 1,0$ are derived from the phase shift analysis of experimental data, which are usually the differential cross sections of the p$^9$Be elastic scattering. The pure phase shifts with the isospin $T = 1$ are obtained from the phase shift analysis of the p$^9$Be or n$^9$Be elastic scattering. Consequently, one can find pure $T = 0$ phase shifts for the p$^9$Be scattering and construct the interaction potential which should correspond to the potential of the BS of the p$^9$Be system in $^{10}$Be.[78] Just the same method of splitting of phase shifts and potentials was used for the p$^3$H system[33,101] and has demonstrated its total availability.

However, we have not found the data on phase shifts for the n$^9$Be, p$^9$B and p$^9$Be elastic scattering at astrophysical energies,[102] therefore, here we will consider only isospin-mixed potentials of the scattering processes in the p$^9$Be system and pure potentials for BSs with $T = 0$, which, as earlier, are constructed on the basis of the description of the BS characteristics – binding energy, charge radius and AC. Exactly the same approach we used earlier for the p$^6$Li and p$^7$Li systems and the potential is selected in a simple Gaussian form with a point-like Coulomb term (2).

Since we don't have the phases of the p$^9$Be elastic scattering obtained from the phase shift analysis of experimental data, we will only rely on the purely qualitative views about their behavior as an energy function. It is known, in particular, that there is $J = 1^-$ over threshold level with $T = 0+1$, energy 0.319(5) MeV (l.s.) and 133 keV (l.s.) width.[78,103] This resonance state can be formed by the $^3S_1$ configuration in the p$^9$Be channel of $^{10}$B because $J(^9Be) = 3/2^-$ and $J(p) = 1/2^+$. The presence of such a level leads to the resonance of the phase shift which equals 90° at this energy.

However, the resonance of the *S*-factor measured in Ref. 28 is observed at the energy of 299 keV (l.s.) which is listed in Table 1 and Fig. 4 of Ref. 28. At the same time the value of 0.380(30) MeV (l.s.) with the width of 330(30) keV (l.s.) is given for the resonance energy in Table 2 of Ref. 28. Both of these values do not correspond to the well-known data.[78,103] That is why an additional analysis of the experimental results was carried out in later[104,105] and the values of 328–329 keV (l.s.) with the width of 155–161 keV were obtained for the energy of this level, which slightly differs from data of Refs. 78 and 103.

Since there is a large difference between various data, we slightly varied parameters of this potential for receiving the best description of the *S*-factor resonance location given in Ref. 28. Consequently, the potential of the $^3S_1$-wave of scattering was obtained, which leads to the resonance of the 90° phase shift at 333 keV (l.s.) and has the following parameters

$$V_0 = -69.5 \text{ MeV}, \quad \gamma_J = 0.058 \text{ fm}^{-2}. \qquad (11)$$

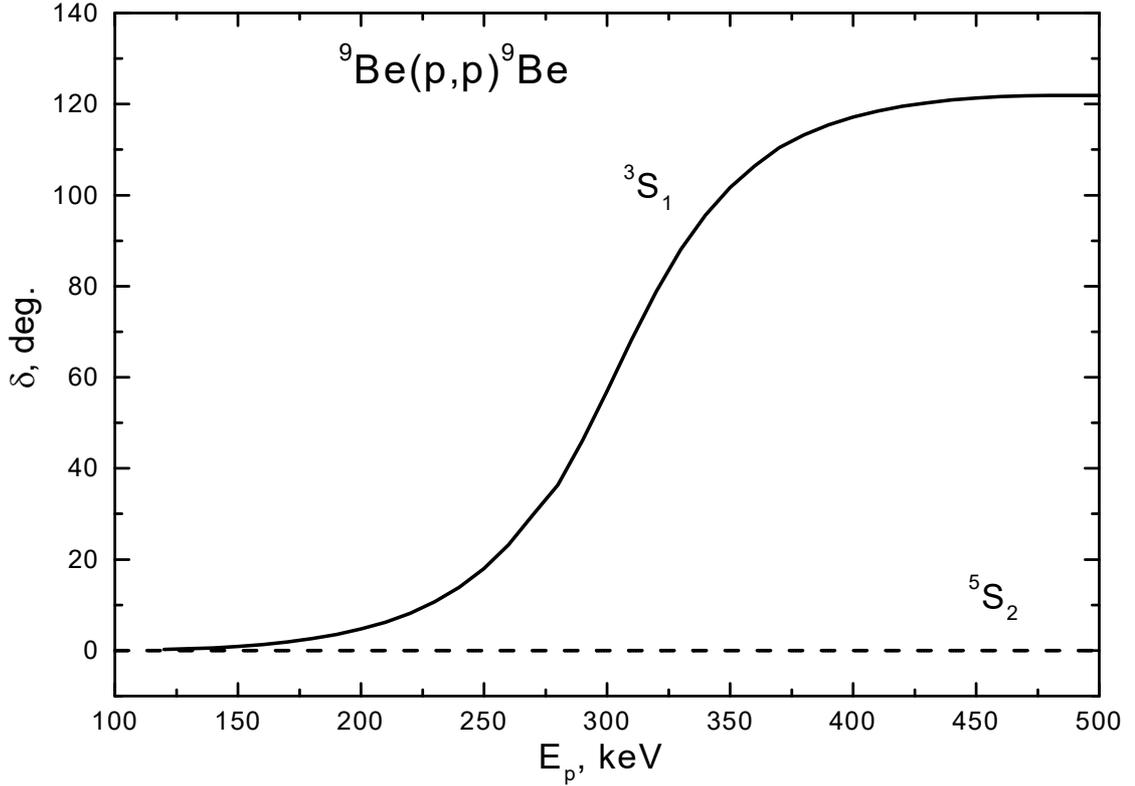

Fig. 7. *S* phase shifts of the elastic p$^9$Be scattering at low energies. Lines – results obtained using Gaussian potentials with parameters that are given in the text.

The triplet $^3S_1$ phase shift of this potential is shown in Fig. 7 by the solid line and has a resonance nature, while the potential itself contains two FSs in accordance with the last variant of the given above classification.

If we use the expression for calculation of the level width using δ phase shift of scattering[106]

$$\Gamma_{l.s.} = 2(d\delta/dE_{l.s.})^{-1}, \qquad (12)$$

then the width of such resonance approximately equals 150(3) keV (l.s.), which is in a complete agreement with the results of Refs. 104,105.

Furthermore, we assume that the $^5S_2$ phase shift almost vanishes to zero at the energy range up to 600 keV which will be considered here because there are no such levels at $^{10}$B spectra comparable to this partial wave at these energies.[78,103] Practically zero phase shift is obtained with the Gaussian potential and parameters

$$V_0 = -283.5 \text{ MeV}, \quad \gamma_J = 0.3 \text{ fm}^{-2}. \qquad (13)$$

It contains two FSs as it follows from the classification of orbital states given earlier. The phase shift of scattering is shown in Fig. 7 by the dashed line. Of course, it is possible to obtain the $^5S_2$ phase shift in the vicinity of zero by using some other variants of potential parameters with two FSs. In this regard it is not possible to fix its parameters definitely and other combinations of $V_0$ and γ are possible. However, additional calculations of the *E*1 transition from the elastic $^5S_2$ wave to the $^5P_3$ BS have shown the weak dependence of the *S*-factor of the proton radiative capture on $^9$Be by the parameters of this potential. Only the zero values of the scattering phase shifts play the main role here.

The following parameters of the potential of the $^5P_3$ BS of the p$^9$Be system corresponding to the GS of $^{10}$B in the cluster channel under consideration were obtained

$$V_0 = -719.565645 \text{ MeV}, \quad \gamma_J = 0.4 \text{ fm}^{-2}. \tag{14}$$

With this potential we have obtained the binding energy of –6.585900 MeV with the accuracy of $10^{-6}$ MeV,[107] the root-mean-square radius is equal to 2.58 fm while the experimental value is 2.58(10) fm[78] and the AC calculated by Whittaker functions is equal to $C_W$ = 2.94(1). Values $R_p$ = 0.8768(69) fm[87] and $R_{Be}$ = 2.519(12) fm[78] were used for cluster radii. The AC $C_W$ error is estimated by its averaging at the range of 5–15 fm where the AC is practically stable. In addition to the allowed BS corresponding to the GS of $^{10}$B such the $P$ potential has two FSs in complete correspondence with the classification of orbital cluster states which was given above.

For the purposes of comparison, we give the results of Ref. 108 for the AC where its value equals $C_W$ = 2.37(2) fm$^{-1/2}$. It is necessary to note, that in this work we use this expression for the determination of the AC

$$\chi_L(R) = C_W W_{-\eta L+1/2}(2k_0 R), \tag{15}$$

which differs from our definition (3) for $\sqrt{2k_0}$ value. To bring these constants to a unified dimensionless form the results should be divided by $\sqrt{2k_0}$, where $k_0$ = 0.536 fm$^{-1}$ for the p$^9$Be system. Then we receive for AC in our definition the value of 2.29, which substantially differs from the above result. However, if we take AC value obtained above, the charge radius of $^{10}$B will be somewhat underestimated because the "tail" of the WF decreases more sharply.

The following parameters for the potentials of the first three excited but not BSs in the p$^9$Be channel with $J^PT$ = $1^+0$, $0^+1$ and $1^+0$ at energies 0.71835, 1.74015 and 2.1543 MeV (see Ref. 78) were received

$$V_0(0.718350) = -715.162918 \text{ MeV}, \quad \gamma_J = 0.4 \text{ fm}^{-2}, \tag{16}$$

$$V_0(1.740150) = -708.661430 \text{ MeV}, \quad \gamma_J = 0.4 \text{ fm}^{-2}, \tag{17}$$

$$V_0(2.154300) = -705.935443 \text{ MeV}, \quad \gamma_J = 0.4 \text{ fm}^{-2}. \tag{18}$$

They describe precisely the values of the given above energy levels and shown, which, relative to the p$^9$Be channel threshold, are equal to –5.867550, –4.845700 and –4.431600 MeV. They properly lead to charge radii of 2.59, 2.60 and 2.61 fm, ACs of 2.74(1), 2.46(1) and 2.35(1) at the range from 4–5 fm to 11–13 fm and have two FSs and one AS.

The variational method with the expansion of the cluster WF of the p$^9$Be system in non-orthogonal Gaussian basis $\Phi_L(R) = \dfrac{\chi_L(R)}{R} = R^L \sum_i C_i \exp(-\beta_i R^2)$ is used for an additional control of the accuracy of the binding energy calculations.[106] The GS energy equals –6.585896 MeV was obtained for this potential with the order of matrix $N$ = 10 which differs from the above finite-difference value[107] by 4 eV only. Residuals have $10^{-11}$ order, AC at the range 5–10 fm equals 2.95(3), and the charge radius does not differ from the previous results. Expansion parameters of the received variational GS radial WF of $^{10}$B in the p$^9$Be cluster channel are listed above in table in Ref. 107.

As we repeatedly noted that the variational energy decreases as the dimension of the basis increases and gives the upper limit of the true binding energy, but the finite-difference energy increases as the size of steps decreases and the number of steps increases.[107] Therefore, it is possible to use the average value of –6.585898(2) MeV for the real binding energy in this potential. Meanwhile, the calculation error of finding the binding energy of $^{10}$B in the p$^9$Be cluster channel using two different methods is about ±2 eV.

In the frame of the VM we obtained that the value of energy equals –4.845692 MeV, the charge radius of 2.61 fm and the AC is equal to 2.48(2) in the range 5–12 fm for the potential of the second excited state. The average energy value –4.845696(4) MeV is obtained for this level using two different methods and two different computer programs and residuals are of the order of $10^{-13}$.

### 4.3. *Astrophysical S-factor of the radiative proton capture on $^9$Be*

While considering electromagnetic transitions we will take into account the $E1$ process, denote it as the $E1$(BS), from the resonance $^3S_1$ wave of scattering to three excited, but BSs in the p$^9$Be cluster channel of $^{10}$B with $J^\pi T = 1^+0$, $0^+1$ and $1^+0$ (see Ref. 78) denoting them as the $^3P$ states. As well as the $E1$ transition from the $^5S_2$-wave of scattering with zero phase shift to the ground $^5P_3$ BS of this nucleus denoting it as $E1$ (GS).

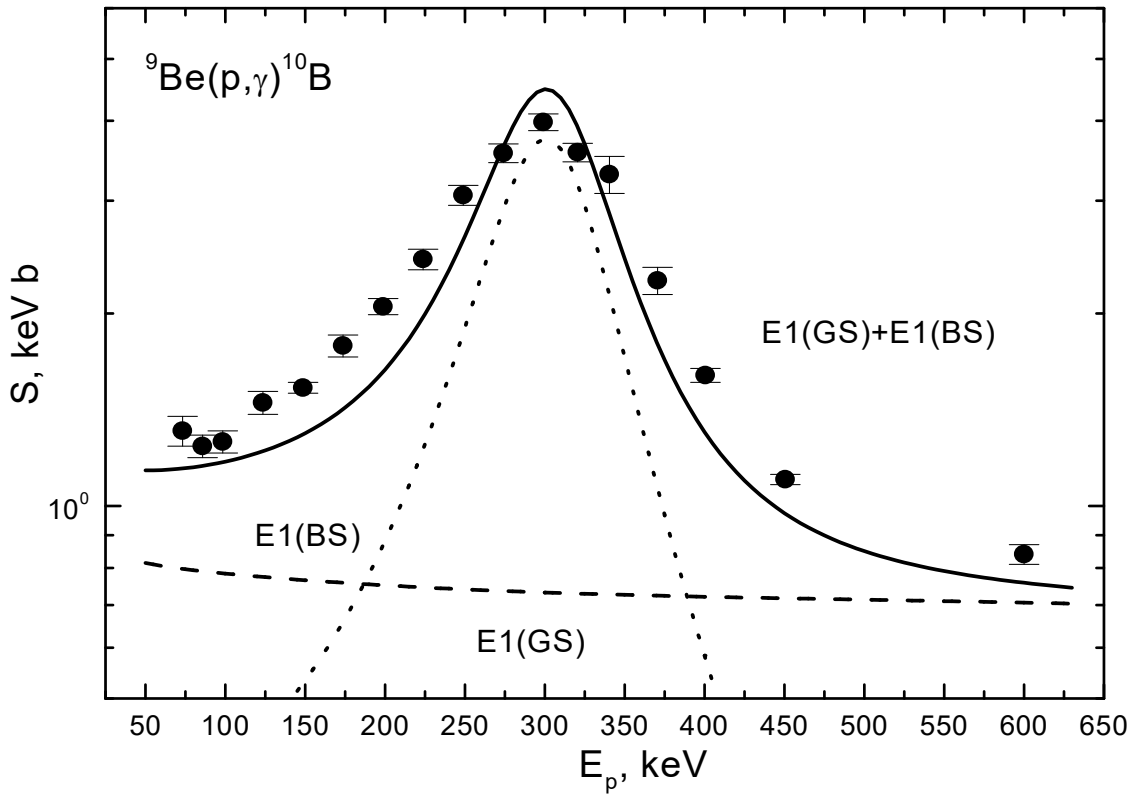

Fig. 8. Astrophysical *S*-factor of the proton radiative capture on $^9$Be reaction. Dots – experimental data from Ref. 28. Lines – calculation results for different electromagnetic transitions with the potentials mentioned in the text.

The calculation results and experimental data for the *S*-factor at the energy range 50÷600 keV (l.s.) from Ref. 28 are shown in Fig. 8 by the solid line. Obviously, the value of the total calculated *S*-factor at the energy range 50÷100 keV remains almost constant and equals 1.15(2) keV b, which is in quite a good agreement with the data of Ref. 28, where the average of the first three experimental points at the energy 70÷100 keV equals 1.27(4) keV b.

The transition to the $^5P_3$ GS of $^{10}$B from the $^5S_2$ scattering wave leads to the value of 0.81 keV b for the calculated *S*-factor at the energy 50 keV (dashed line in Fig. 8). Line extrapolation of the received result to zero energy gives approximately 0.90 keV b. The sum of transitions from the $^3S_1$ scattering wave to the three bound $^3P$ levels is shown in Fig. 8 by the dotted line.

For comparison we will give some extrapolation results to zero energy for different experimental data. For instance, the value 0.92 keV b was obtained in Ref. 109 for the *S*-factor with transition to the GS, which is in complete agreement with the value of 0.90 keV b received above. However, the values 1.4 keV b, 1.4 keV b and 0.47 keV b are given for the transitions to the three excited levels considered above with $J^PT = 1^+0$, $0^+1$ and $1^+0$ respectively[109] and their sum evidently exceeds our result and data from Ref. 28. Furthermore, the value 0.96(2) keV b for the total *S*-factor was received in the later[104] and the value from 0.96(6) to 1.00(6) keV b was found in one of the latest[105] devoted to this reaction. Both of these values are in good agreement with the values obtained above.

## 5.   Neutron-capture reaction $^{10}$Be(n, γ)$^{11}$Be

### 5.1. *Structure of states in the n$^{10}$Be system*

For $^{10}$Be, as well as for $^{10}$B,[38] we accept the Young tableaux {442}, so for the n$^{10}$Be system we have {1} × {442} → {542} + {443} + {4421}.[77,79] The first from the obtained tableaux compatible with orbital angular moments $L = 0, 2, 3, 4$, is forbidden as it is forbidden to have more than four nucleons in the *s*-shell. The second tableau is allowed and compatible with the orbital angular moments $L = 1, 2, 3, 4$; and the third is also allowed, compatible with $L = 1, 2, 3$.[77]

Generally speaking, the lack of the tables of products of the Young tableaux for the number of particles 10 and 11 makes it impossible to make an accurate classification of the cluster states in the given system of particles. However, even such a qualitative estimation of orbital symmetries allows one to determine the presence of the FSs in the *S* wave and lack of the FSs for the *P* states. Exactly such a structure of the FSs and ASs in the different partial waves allows one to further build the potentials of intercluster interaction, which are necessary for the calculation of the total cross sections for considering the radiative capture reaction.[37,38]

Thus, taking into account only the lower partial waves with orbital angular moments $L = 0, 1, 2$, it can be said that for the n$^{10}$Be system (for $^{10}$Be $J^\pi, T = 0^+, 1$; see Ref. 78) in the potentials of the $^2P$ waves only the AS presents, and the $^2S$ and $^2D$ waves have FSs. The state in the $^2S_{1/2}$ wave (in the notation of $^{(2S+1)}L_J$), corresponds to the GS of $^{11}$Be with $J^\pi, T = 1/2^+, 3/2$ and is located at the binding energy of the n$^{10}$B system of -0.5016 MeV.[110]

Note that the $^2P$ waves correspond to the two allowed Young tableaux {443} and {4421}. This situation seems to be similar to the systems N$^2$H or N$^{10}$B, described elsewhere,[37,38,46–49,80] where the potentials for the scattering processes depend on two Young tableaux, and for the BS only on one.[111,112] Therefore, here we assume that the potential $^2P_{1/2}$ of the BS (first excited state – FES) corresponds to one tableau {443}. Consequently, the potentials of the $^2P_{1/2}$ BS and of the $^2P_{1/2}$ scattering processes are different, because they depend on a different set of Young tableaux. To fix the idea, we will assume that for a discrete spectrum the AS in the $^2P_{1/2}$ wave is bound, while for the scattering processes it is not bound. Therefore, the depth of such a potential can be simply set equal to zero. The FS occurs to be the BS for the $^2S_{1/2}$ scattering wave or for the discrete spectrum in the n$^{10}$B system.

Now let us consider the FES, bound in the n$^{10}$Be channel, and the first resonance state (FRS) of $^{11}$Be,[110] which is not bound in the n$^{10}$Be channel and corresponds to the resonance in the n$^{10}$Be scattering. The FES of $^{11}$Be is located at an energy of 0.32004 MeV relative to the GS with the $J^\pi = 1/2$ moment or -0.18156 MeV relative to the n$^{10}$Be channel threshold. This state can be related to the doublet $^2P_{1/2}$ level without FS. The first resonance state is located at 1.783 MeV relative to the GS or at 1.2814 MeV relative to the n$^{10}$Be channel threshold. For this level the moment $J^\pi = 5/2^+$ is provided,[110] which allows one to take $L = 2$ for it, i.e., to consider it as the $^2D_{5/2}$ resonance in the n$^{10}$Be system at 1.41 MeV (l.s.), and its potential has the FS. The width of such resonance is equal to 100(10) keV in the center of mass (c.m.).[110]

On the basis of these data, it is reputed that $E1$ capture from the $^2P$ scattering waves with the potential of zero depth without the FSs to the $^2S_{1/2}$ GS of $^{11}$Be with the bound FS is possible.

$$\text{Process No.1.} \quad \begin{array}{c} ^2P_{1/2} \to {}^2S_{1/2} \\ ^2P_{3/2} \to {}^2S_{1/2} \end{array}.$$

For radiative capture to the FES, the similar $E1$ transition from the $^2S_{1/2}$ and $^2D_{3/2}$ scattering waves with the bound FSs to the $^2P_{1/2}$ BS without FS is possible.

$$\text{Process No.2.} \quad \begin{array}{c} ^2S_{1/2} \to {}^2P_{1/2} \\ ^2D_{3/2} \to {}^2P_{1/2} \end{array}.$$

The GS potentials and the FES will be constructed further in a way for the correct description of the channel binding energy, the charge radius of $^{11}$Be and its AC in the n$^{10}$Be channel. Therefore, the known values of the ANC and the spectroscopic factor $S$, with the help of which the AC is found, have a quite a large error. The GS potentials will also have several variants with different parameters of width, which strongly affect the value of the AC.

The data on the $A_{NC}$, for example, have been given elsewhere.[113] Here we also will use the well-known relation (4) where $S$ – the spectroscopic factor, $C$ – asymptotic constant in fm$^{-1/2}$, which is connected to the dimensionless AC $C_W$,[59] used by us as follows: $C = \sqrt{2k_0} C_W$, and the dimensionless constant $C_W$ is given by the expression (3).[59]

In further calculations we used the radius of $^{10}$Be in the GS equal to 2.357(18) fm,[114] and for the GS of $^{11}$Be the known radius value of 2.463(15) fm.[110] The charge radius of the neutron is assumed to be zero, and its massive range 0.8775(51) fm coincides with the known radius of the proton.[87] In addition, for the charge radius of the FES of $^{11}$Be the calculated value of 2.43(10) fm is known,[115] and for the GS the value 2.42(10) fm is obtained in the same paper. The radius of the neutron in $^{11}$Be estimation is equal to 5.6(6) fm.[115] At the same time, for the neutron radius in the GS the value of 7.60(25) fm is given,[113] while for the FES -4.58(25) fm. In the all calculations for the masses of nucleus and neutron the exact values are used: $m(^{10}\text{Be}) = 10.013533$ amu,[86] and $m(n) = 1.00866491597$ amu.[87]

The total cross sections of the radiative capture $\sigma(NJ,J_f)$ for the *EJ*- and *M*1-transitions in the potential cluster model are given elsewhere.[37,38,46–49,58] For the neutron magnetic moment and $^{10}$Be the following values were used: $\mu_n = -1.91304272\mu_0$ and $\mu(^{10}\text{Be}) = 0$,[78] where $\mu_0$ – nuclear magneton. The correctness of this expression for the *M*1 transition is pre-tested in our previous papers[37,38,42,80] on the basis of radiative neutron capture reaction on $^7$Li and proton capture reaction on $^2$H at low and astrophysical energies.

## 5.2. *The n$^{10}$Be interaction potentials*

As usual,[37,38,42,46–49,80–82] we use the potential of the Gaussian form with the point-like Coulomb term (2) in the capacity of the n$^{10}$Be interaction in each partial wave with a given orbital angular moment *L*.

The GS of $^{11}$Be in the n$^{10}$Be channel is the $^2S_{1/2}$ level and this potential should describe the AC of this channel correctly. In order to extract this constant from the available experimental data, let us consider information about the spectroscopic factor *S* and the $A_{NC}$. Results for $A_{NC}$ of Ref. 113 are presented in Table 5 from paper[116] – with additional results from another study.[39] In addition to this, a relatively large amount of data for the spectroscopic factors of the n$^{10}$Be channel of $^{11}$Be is contained in a further study,[110] so we give their values in the separate Table 6.[116]

Table 5. The $A_{NC}$ data of $^{11}$Be in the n$^{10}$Be channel

| Reaction from which the $A_{NC}$ is determined | The value of the $A_{NC}$ in fm$^{-1/2}$ for the GS | The value of the $A_{NC}$ in fm$^{-1/2}$ for the FES | Reference |
|---|---|---|---|
| (d,p$_0$) at 12 MeV | 0.723(16) | 0.133(4) | 113 |
| (d,p$_0$) at 25 MeV | 0.715(35) | 0.128(6) | 113 |
| (d,p$_0$) at 25 MeV | 0.81(5) | 0.18(1) | 39 |
| | 0.68–0.86 | 0.122–0.19 | *Data Interval* |
| | *0.749* | *0.147* | Average $\overline{A}_{NC}$ |

Furthermore, based on Equation (4) for the GS we find $\overline{A}_{NC}/\sqrt{\overline{S}} = \overline{C} = 0.94$ fm$^{-1/2}$, and since $\sqrt{2k_0} = 0.546$, then the dimensionless AC is defined as $\overline{C}_w = \overline{C}/\sqrt{2k_0}$, is equal to $\overline{C}_w = 1.72$. However, the range of values of the spectroscopic factor is so high that the value $C_w$ may be in the range of 1.54–2.29, and if we consider the errors of $A_{NC}$, then this interval can be extended to 1.40–2.63. For the FES at $\sqrt{2k_0} = 0.423$ we find $\overline{C}_w = 0.45$ similarly, and the range of values $\overline{C}_w$ for the average ANC is equal to 0.35–0.62. If we consider the $A_{NC}$ errors, then this interval is expanded to 0.29–0.81.

Table 6. Data for the spectroscopic factors $S$ of $^{11}$Be in the n$^{10}$Be channel

| The $S$ value for the GS | The $S$ value for the FES | Reference |
|---|---|---|
| 0.42(6) | 0.37(6) | 117 |
| 0.72(4) | --- | 118 |
| 0.61(5) | --- | 119 |
| 0.56(18) | 0.44(8) | 120 and 121 |
| 0.73(6) | 0.63(15) | 122 and 123 |
| 0.77 | 0.96 | 124 |
| *0.36–0.79* | *0.31–0.96* | *Data Interval* |
| 0.64 | 0.6 | Average $\overline{S}$ |

The potential $^2S_{1/2}$ of the GS with the FS, which allows one to obtain the dimensionless constant $C_W$, close to the average value of 1.72, has the parameters[116]

$$V_{1/2} = 47.153189 \text{ MeV and } \gamma_{1/2} = 0.1 \text{ fm}^{-2}. \tag{19}$$

This leads to the binding energy of -0.501600 MeV with an accuracy of the finite-difference method (FDM) for calculating the binding energy of $10^{-6}$ MeV,[80] the AC $C_W = 1.73(1)$ on the interval 7–30 fm, the mass radius of 3.16 fm, the charge radius of 2.46 fm. The determination of the estimated expressions of these radiuses is given elsewhere.[37,38,64] The AC errors are determined by averaging over the specified range of distances.

Such a potential of the GS with the FS is in a full accordance with the classification of states according to Young tableaux given above, and gives the charge radius of $^{11}$Be which agrees well with data elsewhere.[110] The parameters of the GS potential were built on the basis of the exemplary description received above average value of the AC equal to 1.72, and its phase shift is shown in Fig. 9 by the solid line. This potential at the orbital angular moment $L = 2$ leads to the non-resonant $^2D$ scattering phase shift without spin-orbital splitting, shown in Fig. 9 by the dotted line. On the same figure, the $^2S_{1/2}$ phase shifts of the n$^{10}$Be scattering, obtained in the calculations of another study,[125] are shown by the points.

To compare the results, let us consider another variant of the GS potential with the FS and the parameters

$$V_{1/2} = 68.68161 \text{ MeV and } \gamma_{1/2} = 0.15 \text{ fm}^{-2}. \tag{20}$$

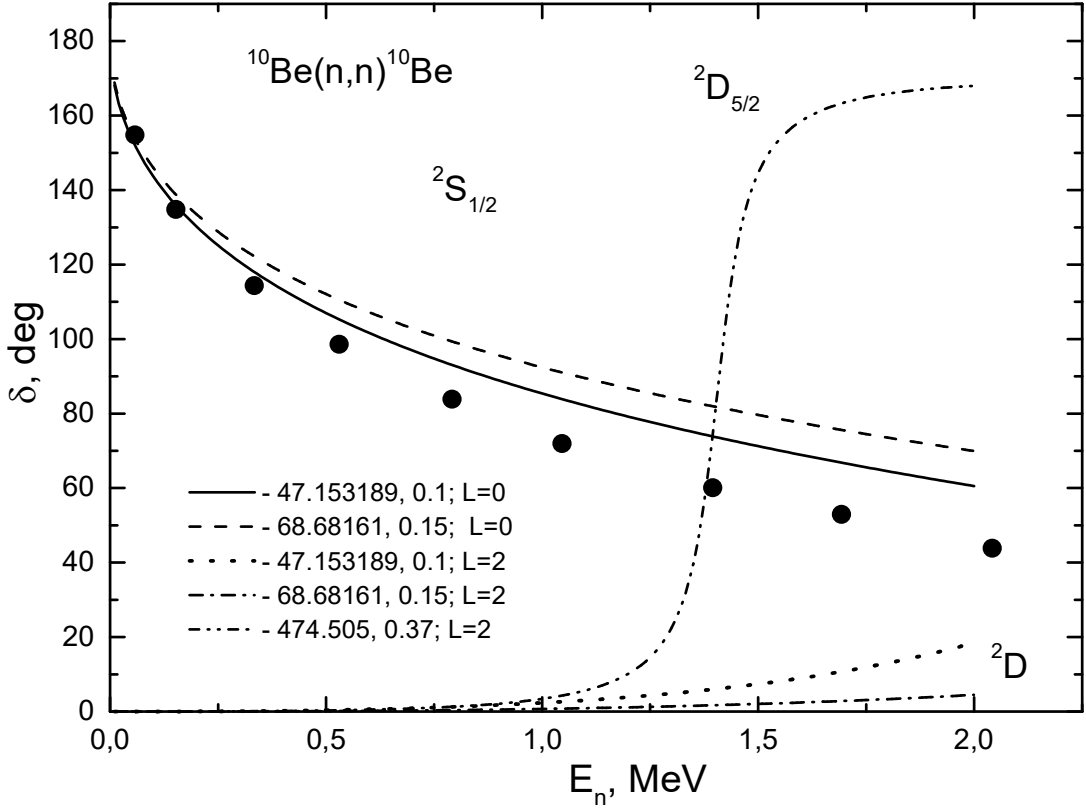

Fig. 9. The n$^{10}$Be elastic scattering phase shifts of the $^2S_{1/2}$ wave. The meaning of the lines and points is explained in the text.

The potential leads to the binding energy of -0.501600 MeV at the same accuracy of the FDM, $C_W = 1.56(1)$ on the interval of 5–30 fm, the mass radius of 3.05 fm, the charge radius of 2.45 fm and its phase shift is shown in Fig. 9 by the dashed line. The dimensionless AC is close to the lower limit of the range of values of this magnitude 1.54–2.29. Such a potential with the orbital angular moment $L = 2$ leads to the $^2D$ scattering phase shift without spin-orbital splitting, shown in Fig. 9 by the dash-dotted line.

Note that for the potential of the resonance $^2D_{5/2}$ wave with the FS, which will be required further for the consideration of the $E2$ transitions, the following parameters were obtained:

$$V_{5/2} = 474.505 \text{ MeV and } \gamma_{5/2} = 0.37 \text{ fm}^{-2}. \qquad (21)$$

The potential leads to a resonance at 1.41 MeV in the l.s. at a width of $\Gamma_{c.m} = 100$ keV, which is fully consistent with other data,[110] and its phase shift is shown in Fig. 1 by the dot-dot-dashed line.

The $^2P_{1/2}$ potential of the FES without the FS can have the parameters

$$V_{1/2} = 9.077594 \text{ MeV and } \gamma_{1/2} = 0.03 \text{ fm}^{-2}. \qquad (22)$$

The potential leads to the binding energy of -0.181560 MeV with an accuracy of the FDM of $10^{-6}$ MeV, the AC equals 0.73(1) on the interval of 11–30 fm, the mass radius of 3.58 fm and the charge radius of 2.52 fm. The phase shift of such a potential decreases gradually and at 2.0 MeV has the value of approximately 115°. The potential parameters of the FES (22) were chosen for the correct description of the total cross sections of the neutron capture on $^{10}$Be at thermal energy of 25.3 meV, obtained elsewhere,[39] and the value of its dimensionless AC is in the allowable range of 0.29–0.81.

Now we return to the criteria of constructing potentials for the $^2P$ scattering wave, which may differ from the potential of the FES because of the different Young tableaux of these states.[111,112] First, as was indicated above, such a potential should not have FS. Therefore, we do not have the results of the phase shift analysis of the n$^{10}$Be elastic scattering, and in the spectra of $^{11}$Be at energies below 2.0 MeV there are no resonances of the negative parity; we will consider that the $^2P$ potentials should lead in this energy range to almost zero scattering phase shifts – so they simply may have the null depth. For the potential of the $^2S_{1/2}$ scattering, the interaction $^2S_{1/2}$ of the GS with the FS will be used, for instance, the variant of the potential (19), because it leads to a relatively good agreement with the scattering phase shifts,[125] shown by the solid line and the points in Fig. 9.

### 5.3. *Total cross sections for the neutron radiative capture on $^{10}$Be*

As mentioned, we assume that the $E1$ radiation capture No. 1 comes from the $^2P$ scattering wave to the $^2S_{1/2}$ GS of $^{11}$Be in the n$^{10}$Be channel. Our calculations of such capture cross sections for the potential of the GS (19) lead to the results that are shown in Fig. 10 by the dashed line,[116] and the results for the potential of the GS (20) presented by the solid line. In all of these calculations for the $^2P$ elastic scattering potentials, the potential of zero depth was used. The experimental data of the neutron radiative capture on $^{10}$Be are shown in Fig. 10 by points and are given in one study[126] with reference to another.[127] As seen from these results, the calculated cross sections describe the available experimental data in a relatively narrow energy range, approximately from 0.3–0.4 MeV to 2.0 MeV. The calculated line decreases rapidly at lower energies and does not describe the data for thermal energy, as shown in Fig. 11. The solid red line in Fig. 10 shows the calculation results from Ref. 126.

Therefore, we will consider the $E1$ transitions to the $^2P_{1/2}$ FES from the $^2S_{1/2}$ and $^2D_{3/2}$ scattering wave further. The results for the $E1$ transition to the GS with the potential (19) and the $^2P$ scattering potential are still shown in Fig. 11 by the dashed line.[116] The cross sections for the $E1$ transition No. 2 from the $^2S_{1/2}+^2D_{3/2}$ scattering wave with the potential (19) for $L = 0$ and 2 to the $^2P_{1/2}$ FES with the potential (22) are shown by the dotted line. The solid line shows the summarized cross section of these two $E1$ processes, which, in general, reproduces the general course of the available experimental data correctly in the given energy region – from the thermal 25.3 meV to 2.0 MeV.

The contribution of the transition from the $^2S_{1/2}$ wave is shown by the dot-dashed line and from the $^2D_{3/2}$ scattering wave by the dot-dot-dashed line. All coefficients in the expression for cross sections were calculated for the $^2D_{3/2}$ wave, and the $^2S_{1/2}$ wave (19) used in these calculations has the orbital angular moment $L = 2$, i.e., it corresponds to the $^2D$ state without the spin-orbital. The calculated cross section at thermal energy was found to be 272 μb. The experimental data at thermal energy were taken from another study[39] – triangle with the value of 290(90) μb and a further study[128] – the square, which indicates the upper limit of the thermal cross section equal to 1 mb.

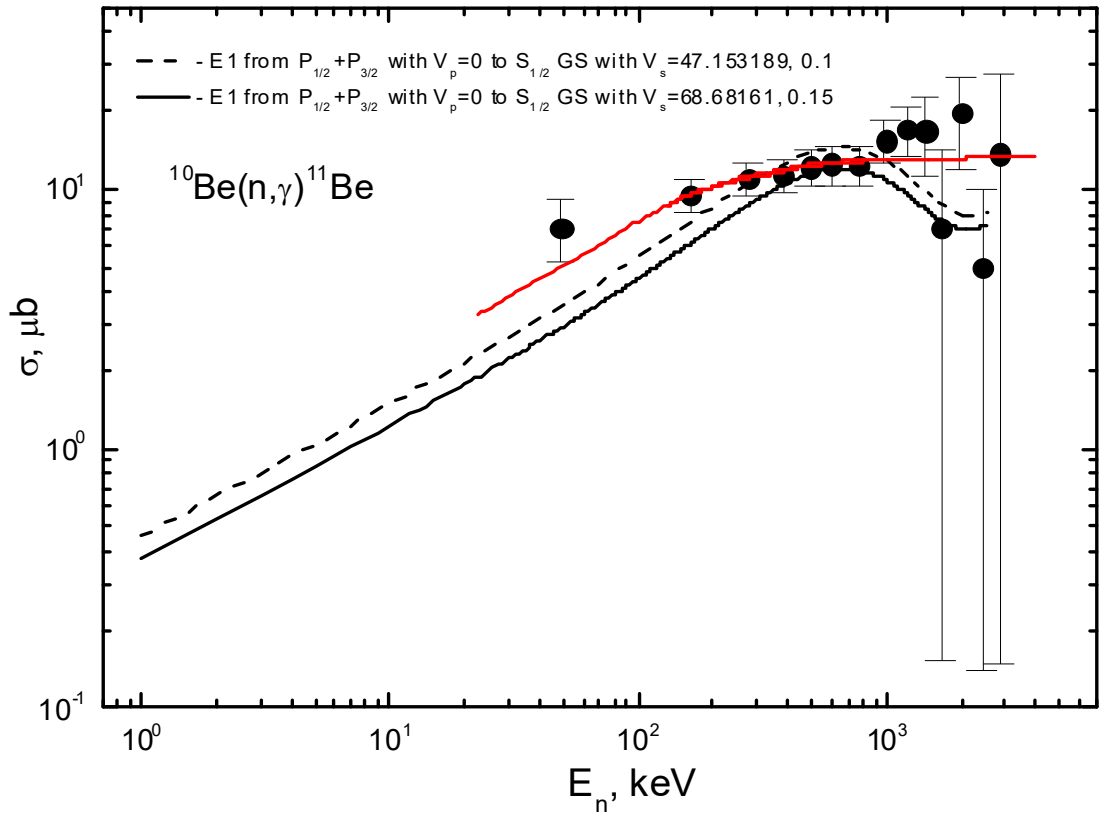

Fig. 10. The total cross sections of the radiative $^{10}$Be(n,γ)$^{11}$Be $E$1 capture to the GS. Points are the experimental data from elsewhere.[126] The meaning of lines is explained in the text.

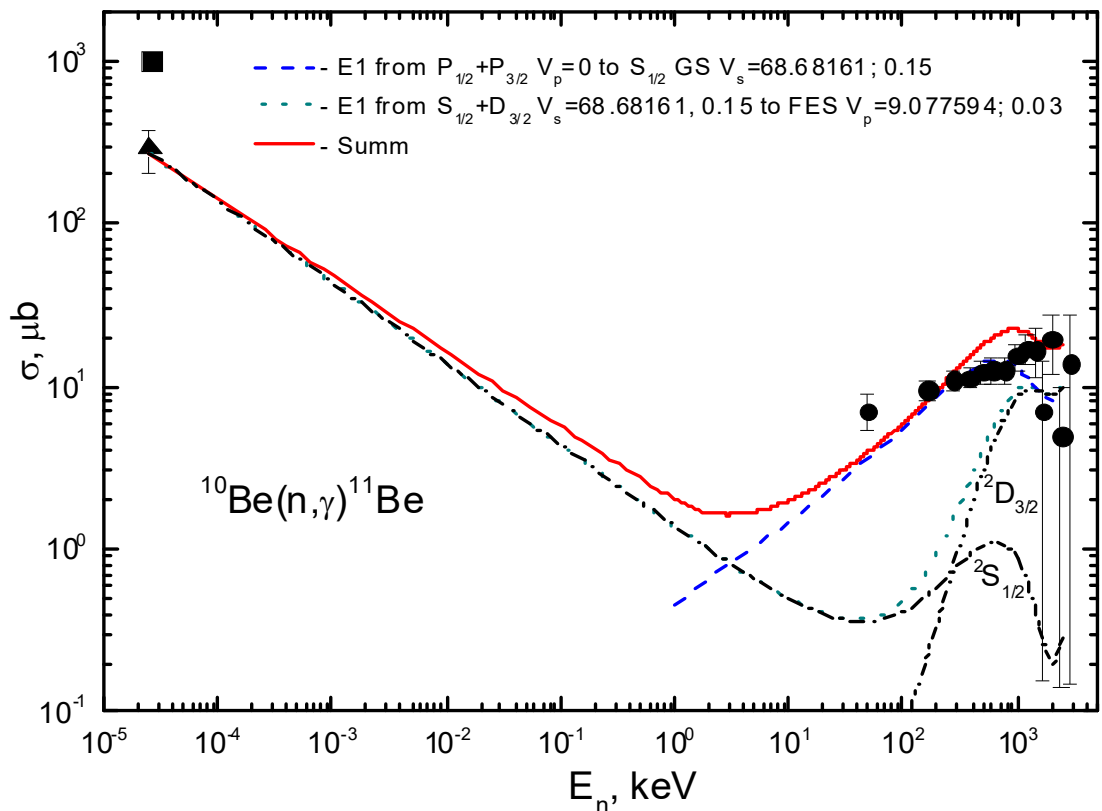

Fig. 11. The total cross sections of the radiative $^{10}$Be(n,γ)$^{11}$Be capture. Experimental data Ref. 127 – points, Ref. 39 – triangle, Ref. 128 – square. The meaning of the lines is explained in the text.

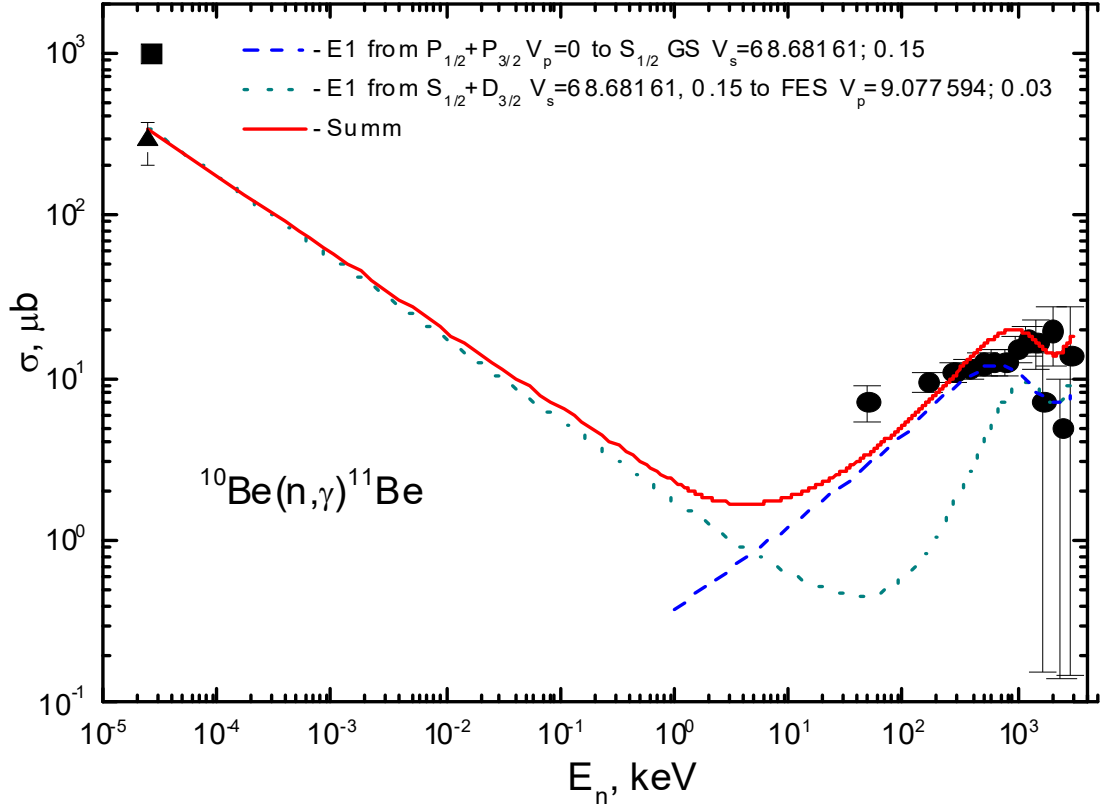

Fig. 12. The total cross sections of the radiative $^{10}$Be(n,γ)$^{11}$Be capture. Experimental data Ref. 127 – points, Ref. 39 – triangle, Ref. 128 – square. The meaning of the lines is explained in the text.

For comparison, now we use the potential of the GS with parameters (20). These results with the same potential of the FES (22) are shown in Fig. 12 (notation as in Fig. 11), with a cross section at thermal energy equal to 343 μb. It is evident that such a potential of the GS slightly better describes the cross section at energies from 0.1–0.2 to 2.0 MeV. Thus, the variants of the calculations with the potentials of the GS (19) and (20) and the potentials of the FES (22) lead to the general description of the available data at the energy of 25.3 meV and in the range of about 0.1–2.0 MeV. The 50–100 keV energy range is described comparatively badly, so we now consider the other variants of the transitions at the neutron capture on $^{10}$Be.

The cross section of the possible $M$1 transition from the $^2S_{1/2}$ scattering wave to the $^2S_{1/2}$ GS of $^{11}$Be in the n$^{10}$Be channel with the same potential (19) or (20) in both states will tend to zero because of the orthogonality of the WFs of discrete and continuous spectra in the same potential. The actual numerical calculation of these cross sections leads to a value of less than $10^{-2}$ μb in the energy range from 1 keV to 2.0 MeV, and at the energy of 25.3 meV the cross section appears to be slightly less than 1% from the cross section of the transition to the FES, shown in Fig. 12 by the dotted line.

If we consider the $M$1 transitions from the $^2P$ scattering waves with the zero potential to the $^2P_{1/2}$ FES with the potential (22), then the sections do not exceed 0.15 μb in the entire energy region. For the $E$2 transitions from the $^2D_{3/2}$ wave with the potential (19) or (20) at $L = 2$ and the $^2D_{5/2}$ wave with the potential (21) to the GS with the $^2S_{1/2}$ even at the resonance energies, the value of these cross sections does not exceed $10^{-3}$ μb. Hence, it is clear that such transitions do not contribute significantly to the total cross section of the considered process; the problem of describing the cross sections in the range from 50 keV to 100–200 keV remains open.

Since at energies of 25.3 meV and up to about 10 eV, the calculated cross section is a straight line (solid line in Fig. 12), it can be approximated by a simple function of energy of the form of Equation (9).

The value of the constant $A = 1.7265$ μb·keV$^{1/2}$ was determined by a single point in the calculated cross section at minimum energy equal to 25.3 meV. The module of the relative deviation (10) of the calculated theoretical cross section ($\sigma_{theor}$) and approximation ($\sigma_{ap}$) of this cross section of Equation (9) in the energy range up to 10 eV is located at the level of 0.1%.

It is realistic to assume that this form of dependence of the total cross section from energy will also be saved at lower energies. Therefore, based on Equation (9), for the approximation of the cross section, one can perform the estimation of the cross section, for instance, at an energy of 1 μeV (1 μeV = $10^{-9}$ keV), which gives a value of about 54.6 mb.

In Fig. 13 the reaction rate $N_A\langle\sigma v\rangle$ of the neutron capture on $^{10}$Be is shown (solid blue line). This corresponds to the solid red line in Fig. 12 and is presented in the form (8),[58] where $N_A\langle\sigma v\rangle$ is the reaction rate in cm$^3$mole$^{-1}$sec$^{-1}$, $E$ is in MeV, the cross section $\sigma(E)$ is measured in μb, μ is the reduced mass in amu and $T_9$ is the temperature in units of $10^9$ K which matches our calculation range of 0.03 to 3.0 $T_9$. Integration of the cross sections was carried out in the range 25.3 meV – 3.0 MeV for 3000 steps with a step value of 1 keV. Green solid line shows the rate for capture reaction to the GS, violet line is the rate for capture reaction to the FES and blue line shows the total reaction rate of the neutron capture on $^{10}$Be.

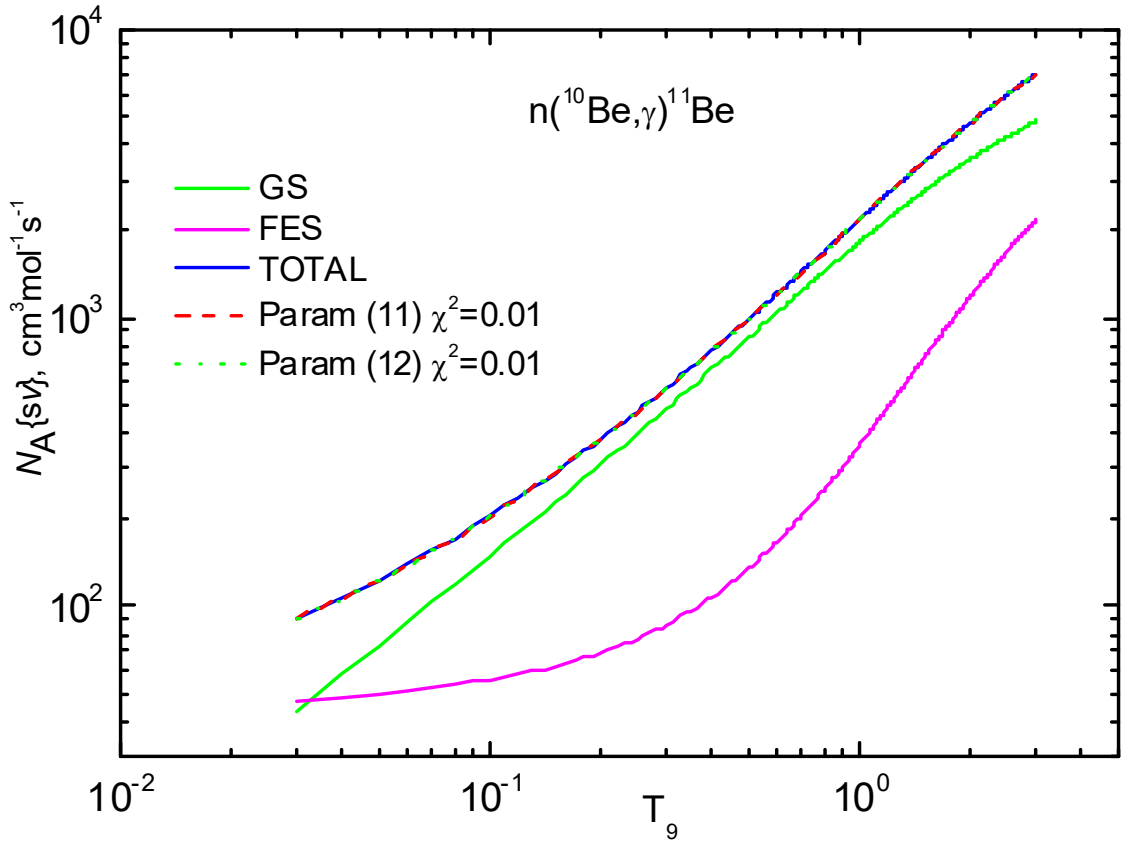

Fig. 13. Reaction rate of the neutron radiative capture on $^{10}$Be. Line is the calculation with potentials which parameters are given in the text.

The resulting shape of the reaction rate in the range of 0.03–3.0 $T_9$ can be approximated by a general polynomial

$$N_A \langle \sigma v \rangle = \sum_{k=1}^{5} a_k T_9^{k-1} \qquad (23)$$

with parameters given in Table 7.

Table 7. Expansion parameters for the reaction rate of the form (23)

| k | 1 | 2 | 3 | 4 | 5 |
|---|---|---|---|---|---|
| $a_k$ | 44.73767 | 1515.535 | 896.4131 | -287.9563 | 25.9271 |

The result of the rate calculation with such parameters is shown in Fig. 13 by the dashed red line at the average value $\chi^2 = 0.01$ at 1% errors of the reaction rate calculated from (8). Increasing the series dimension up to 7 leads to a negligible improvement in the fit of the theoretical curve. However the reduction of dimension to 4 leads to the marked increase of $\chi^2$.

It is possible to use another form of parametrization of the reaction rate (see, for example, Ref. 129)

$$N_A \langle \sigma v \rangle = 930.8611 / T_9^{2/3} \cdot \exp(-1.02168 / T_9^{1/3}) \cdot (1.0 - 6.4647 \cdot T_9^{1/3} + 16.8969 \cdot T_9^{2/3} - 18.0331 \cdot T_9) + (7589.44 \cdot T_9^{4/3} - 3177.42 \cdot 10^4 \cdot T_9^{5/3}) \qquad (24)$$

with $T_9 = 10^9$ K, which also leads to the $\chi^2 = 0.01$ at 1% errors. These results are shown in Fig. 13 by the green dotted line.

## 6. Proton-capture reaction $^{10}$B(p, γ)$^{11}$C

### 6.1. *Classification of the cluster states and the level structure of the p$^{10}$B system*

Furthermore we will assume that it will be possible to accept the orbital Young tableau in the form {442} for $^{10}$B. Therefore, we have {1} × {442} = {542} + {443} + {4421}.[77] The first from the obtained tableaux is compatible with the orbital moments $L = 0,2,3,4$ and is forbidden, because in the nuclear s-shell should not be five nucleons. The second tableau is allowed and is compatible with the orbital moment $L = 1,2,3,4$, and the third, is also allowed and is compatible with $L = 1,2,3$.[77]

As it was said, the absence of product tables of Young tableaux for particle numbers 10 and 11 makes impossible the accurate classification of cluster states in the considered system of particles. However, even so qualitative estimation of the orbital symmetries allows one to determine existence of FSs in the S and D waves and the absence of FSs for the P waves. Exactly this structure of the FSs and the ASs in different partial waves will be used further for construction of the intercluster interaction potentials that are necessary for calculation of the total cross sections of the considered radiative capture reaction at low energies.

Thus, limiting only according to the lowest partial waves with the orbital moment $L = 0,1$ it can be said that only the AS is in the potentials of the $P$ waves for the p$^{10}$B system, but in the $S$ waves there is only FS. Meanwhile, the $P$ wave corresponds to two allowed Young tableaux {443} and {4421}. This situation is analogous, evidently, to the N$^2$H system that was described in the paragraph 3.1, when potentials for the scattering processes depend from two Young tableaux, but for BS only from one. Therefore, here we will consider that the GS potential corresponds to only one tableau {443} – it determines the lowest allowed level for BS in the given partial potential with $L = 1$.[31] Consequently, the BS potentials and the scattering potentials are different, because they depend on the other sets of Young tableaux.[36] Therefore, to fix the idea, we will consider that for the discrete spectrum the ASs in the $P$ waves, which correspond to the GS and to the BS of $^{11}$C, are the bound, but for the scattering processes they are not bound. The FS for the $S$ scattering waves in the p$^{10}$B system is the BS.

Note that in the previous chapter the orbital Young tableau in the form {4411} was obtained for $^{10}$B. Therefore, for the p$^{10}$B system, limiting only by the nuclei of 1$p$ shell, we have {1} × {4411} = {5411} + {4421}.[77] The first from the obtained tableaux compatible with the orbital moments $L = 1,3$ and is forbidden because in the nuclear $s$ shell could not be five nucleons. The second one is allowed and compatible with the orbital moments $L = 1,2,3$.[77] This implies that limiting the orbital moments $L = 0,1$ it can be said that for the p$^{10}$B system in the potentials of the $P$ waves there is the FS with tableau {5411} and the AS with tableau {4421}, which corresponds to the GS of $^{11}$C, and in the $S$ waves the BSs are absent. The results for potentials for the capture total cross sections of this variant of classification will be considered in future, but already in the next chapter for the p$^{11}$B system the Young tableaux {443} and {4421} were used for the GS of $^{11}$B.

Revert to the first variant of classification let us note that the bound allowed p$^{10}$B state in the $^6P_{3/2}$ wave corresponds to the GS of $^{11}$C with $J^\pi, T = 3/2^-, 1/2$ and {443} and locate at the binding energy for the p$^{10}$B system of -8.6894 MeV[110] (it will be recalled that for $^{10}$B $J^\pi, T = 3^+, 0$; see Ref. 110). Some of the p$^{10}$B scattering states and BSs can be mixed by spin with $S = 5/2$ (2$S$+1 = 6) and $S = 7/2$ (2$S$ + 1 = 8), but so long as here we consider only transitions to the $^6P_{3/2}$ GS, then in the future calculations only partial waves at spin $S = 5/2$ will be used.

Let us consider the whole spectrum of resonance levels of $^{11}$C in the p$^{10}$B channel – at the energy lower 1.0 MeV it has three states (see, for example, Table 11.41 in Ref. 110):

1. The resonance at the energy of 0.010(2) MeV (l.s.) with the moment $J = 5/2^+$ and the width of 16(1) keV (l.s.). It corresponds to the level of $^{11}$C at 8.699(10) MeV.

2. The state at the positive energy 0.56(6) MeV (l.s.) with the moment $J = 5/2^+$ and the width of 550(100) keV (l.s.). – it is the level of $^{11}$C at the energy of 9.200(50) MeV.

3. The third resonance at 1.050(60) MeV (l.s.) with the moment $J = 3/2^-$ and the width of 230(50) keV (l.s.). – it is the level of $^{11}$C at the energy of 9.640(50) MeV.

The first and the second from these resonances can be the $^6S_{5/2}$ state of the p$^{10}$B system, and the third is the $^6P_{3/2}$ resonance.

At the low energies the transitions are possible generally from the $S$ scattering waves, therefore at the considering of the $E$1 transitions they are possible only to the $P$

BSs. For example, such transitions are possible to the $^6P_{3/2}$ GS. In addition, the $M$1 processes to the GS from the $P$ waves of the continuous spectrum are probable. Particularly, the $E$1 transition from the $^6S_{5/2}$ scattering wave to the $^6P_{3/2}$ GS is possible:

$$\text{Process No.1. } ^6S_{5/2} \to\, ^6P^1_{3/2}.$$

In addition, the $M$1 transition from two $^6P_{3/2}$ and $^6P_{5/2}$ scattering waves to the GS of $^{11}$C:

$$\text{Process No.2. } ^6P_{3/2} +\, ^6P_{5/2} \to\, ^6P_{3/2}.$$

The first of these waves have the resonance at 1.05 MeV and leads to the resonance transition, but as it can be seen from the results of Ref. 58 where the total cross sections of the proton radiative capture on $^{10}$B are given, its contribution is too small. In addition, on the basis of the form of the $S$-factor of the proton capture on $^{10}$B the resonance at 0.56 MeV[58] in the $^6S_{5/2}$ wave is also practically invisible due to its big width, which is equal, per se, to the energy of this resonance. Thus, besides the $E$1 transition from the $^6S_{5/2}$ scattering wave to the GS, the $M$1 transition from the $^6P_{3/2}$ and $^6P_{5/2}$ scattering waves to the GS will be considered, meanwhile the resonance at 1.05 MeV does not take into account.

### 6.2. *Construction of the p$^{10}$B interaction potentials*

We will use the Gaussian form of interaction (2) with the point-like Coulomb term for construction of the central intercluster potentials. Since, furthermore in the first place the transitions from certain scattering waves to the $^6P_{3/2}$ GS of $^{11}$C in the p$^{10}$B channel will be considered, we will obtain firstly the potential of this state. As it was said the GS lies at the binding energy of -8.6894 MeV in the p$^{10}$B channel and has the moment 3/2$^-$, that is pure by spin $S = 5/2$ the $^6P_{3/2}$ level.[110] Since, we have not find data on the charge radius $^{11}$C, it will be reasonable to consider that there is not big difference between it and the radius of $^{11}$B equals of 2.43(11) fm.[110] The radius of $^{10}$B is known and is equal to 2.4277(499) fm,[102] and the proton radius has the value of 0.8775(51) fm.[87]

Consequently, the parameters of the $^6P_{3/2}$ GS without FS were obtained, as it is followed from the given above classification of FSs and ASs according to Young tableaux for the p$^{10}$B system

$$V_{\text{g.s.}} = -337.1459 \text{ MeV}, \quad \gamma_{\text{g.s.}} = 1.0 \text{ fm}^{-2}. \tag{25}$$

It leads to the charge radius of 2.32 fm, the binding energy of -8.6894 MeV at accuracy of $10^{-4}$ MeV,[130] and the AC equals 1.16 at the range of 2–10 fm. The scattering phase shift of this potential smoothly drops from 180º at zero energy to 179º at 1.0 MeV. As usual, we use the dimensionless AC value that is determined through the Whittaker function (3). In Ref. 131 the value 8.9(8) fm$^{-1}$ is given for square of the dimensional AC of the GS, which has the factor "6" connected with the permutation of nucleons.[132] So as for the dimensional AC we obtain 1.22(5) fm$^{-1/2}$. Since $\sqrt{2k_0} = 1.11$, then for the dimensionless AC we have 1.10(5) value that is in a good agreement with the obtained above value.

Furthermore, it will be possible to construct the potential for the $^6S_{5/2}$ resonance, which lies at 0.010(2) MeV with the moment $J = 5/2^+$ at the width 16(1) keV and corresponds to the level 8.699 MeV of $^{11}$C.[110] This potential can have the parameters

$$V_{S5/2} = -49.8 \text{ MeV}, \quad \gamma_{S5/2} = 0.088 \text{ fm}^{-2}. \tag{26}$$

It contains the FS and, as it will be seen further, reproduces the resonance location in the total cross sections of the proton radiative capture on $^{10}$B Refs. 133,134 at lowest energies practically correct.

The next $^6S_{5/2}$ potential for the resonance at 0.56(60) MeV (l.s.) with the moment $J = 5/2^+$ and the width 550(100) keV can have the parameters

$$V_{S5/2} = -18.293 \text{ MeV}, \quad \gamma_{S5/2} = 0.033 \text{ fm}^{-2}, \tag{27}$$

has the bound FS and correctly reproduce the resonance energy of 560(1) keV at the width of 570(5) keV, and its phase shift is shown in Fig. 14 by the solid line. The scattering phase shift value at the resonance energy is equal to 90(5)°.

It is possible to give the potential parameters without FSs for the $^6P_{3/2}$ resonance at 1.05(6) MeV with the width of 230(50) keV, though we shall not use it in future calculations for a while

$$V_{P3/2} = -31.5 \text{ MeV}, \quad \gamma_{P3/2} = 0.1 \text{ fm}^{-2}. \tag{28}$$

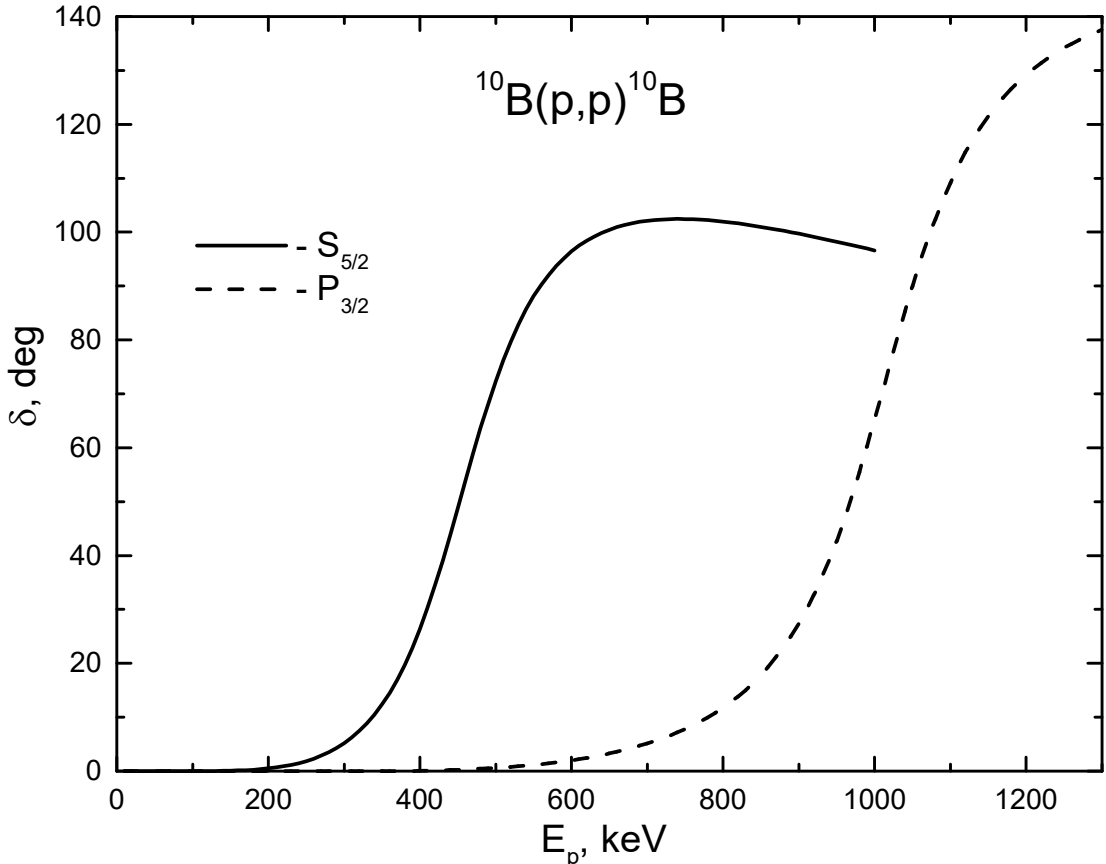

Fig. 14. The $S_{5/2}$ and $P_{3/2}$ phase shifts of the p$^{10}$B elastic scattering at low energies, obtained with the potentials given in the text.

This potential leads to the resonance at 1050(1) keV with the width of 250(5) keV and with the phase shift value of 90(1)°, and its phase shift is shown in Fig. 14 by the dashed line.

Since, in the region up to 1.0 MeV that we consider, the $^6P_{3/2}$ and $^6P_{5/2}$ scattering waves have not resonances, we will consider their phase shifts close to zero, and because they does not contain FSs, then the depth of such potentials, for the first variant, can be equalized to zero. In the capacity of the second variant it is possible to consider that this potential has to have parameters leading to the scattering phase shifts that close, but not exactly equals zero.

### 6.3. *Total cross sections of the radiative proton capture on $^{10}B$*

We would remind one that the experimental data for the total cross sections the astrophysical *S*-factors of the radiative proton capture on $^{10}B$ are given in Refs. 133 and 134. Meanwhile, the numerical values of the *S*-factor, obtained in Ref. 134, were taken from the data base EXFOR, given in the MSU site,[102] and the numerical values of the *S*-factor for the transition to the GS, obtained in Ref. 133 were taken from figures of this work.

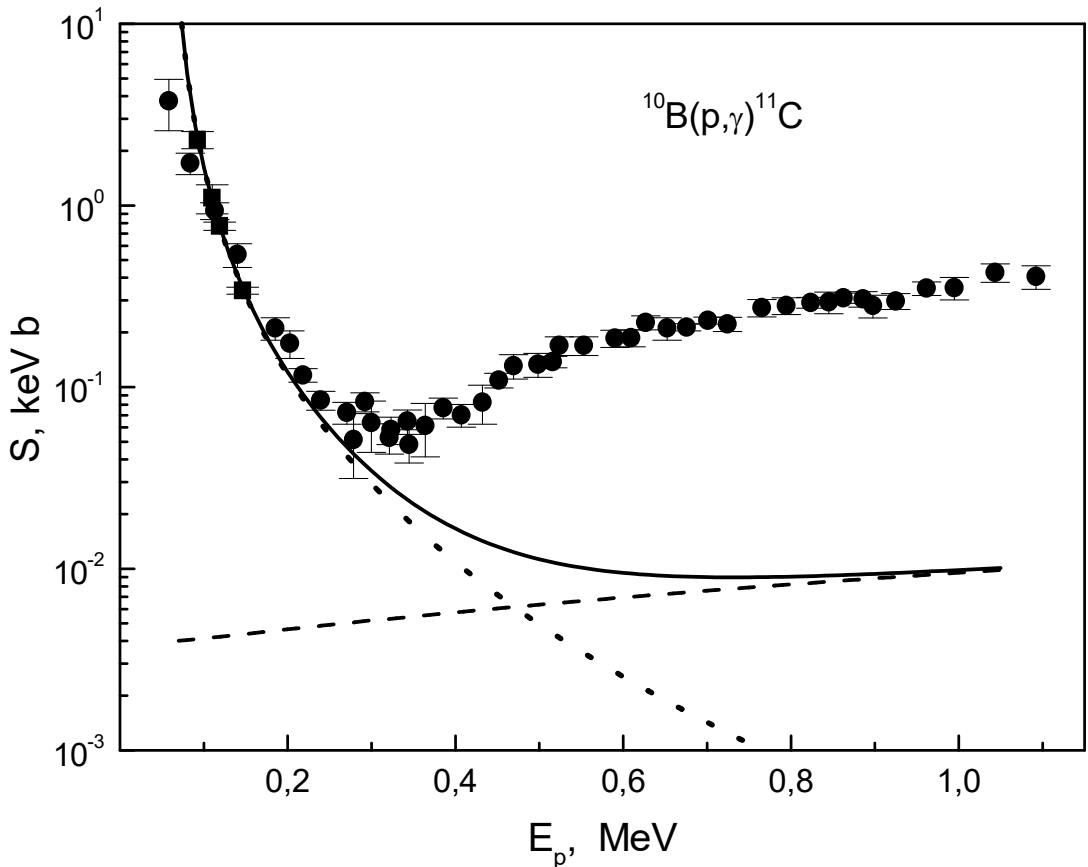

Fig. 15. Astrophysical *S*-factor of the radiative proton capture on $^{10}B$ to the GS. Experiment: points (●) – numerical data on the *S*-factor are from figures of Ref. 133, squares (■) – numerical data on the *S*-factor obtained in Ref. 134 – were taken from data base EXFOR, given in the MSU site.[102]

Consider now the astrophysical *S*-factor of the radiative proton capture on $^{10}B$ to the GS of $^{11}C$ with the potential (25), which shape is shown in Fig. 15 by points and squares.[133,134] The astrophysical *S*-factor does not contain evident resonances at energies 560 keV and 1050 keV, where the resonance states are observed in the elastic p$^{10}$B scattering and resonances in spectra of $^{11}$C.[110] There is only the resonance in the range of zero energy, which corresponds to the resonance in the $^6S_{5/2}$ scattering wave at 10 keV. Therefore the potential of this state was constructed only on the basis of the correct description of the resonance in the capture $S$-factor. Consequently, the potential parameters (26) were obtained, and the form of the calculated $S$-factor of the radiative proton capture on $^{10}$B for the $E$1 transition from the $^6S_{5/2}$ scattering wave to the $^6P_{3/2}$ GS, i.e., the process No. 1 from Sec. 8.1 is given in Fig. 15 and Fig. 16 by the dotted line at low energies.

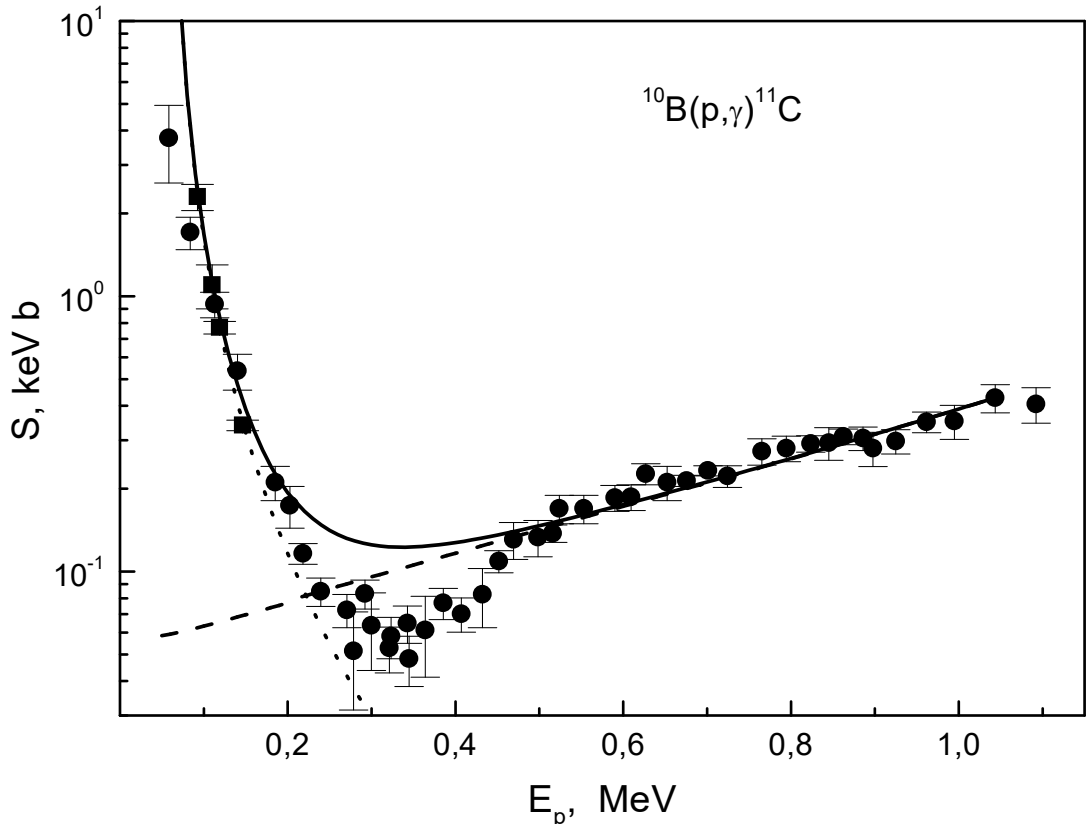

Fig. 16. Astrophysical $S$-factor of the radiative proton capture on $^{10}$B to the GS. Points and squares – experiment as in Fig. 15.

As it is seen from Fig. 15 the calculated $S$-factor is acceptable reproduce the results of two experimental measurements from Refs. 133 and 134 in the range of resonance at 10 keV and to the energy, approximately, 0.25–0.30 MeV. Because, the experimental $S$-factor above 300 keV has not the resonance character, we have considered furthermore only the nonresonance $M$1 transitions from the $^6P_{3/2}$ and $^6P_{5/2}$ scattering waves to the $^6P_{3/2}$ GS – the calculated results are shown in Fig. 15 by the dashed line. The first variant of the potential with the zero depth was used for both scattering waves, because such states haven't resonances at low energies and FSs, and therefore the potential should lead to the phase shifts that close or equal to zero.

The summed cross section of two considered above processes that describes the $S$-factor only in the energy range 0.25–0.30 MeV is shown by the solid line in Fig. 15, meanwhile only at the expense of the resonance transition to the GS at 10 keV. We have not considered here the resonance processes like $M$1 from the $^6P_{3/2}$ scattering wave at 1.05 MeV or the $E$1 transition from the resonance $^6S_{5/2}$ wave at 560 keV to the GS of $^{11}$C because they lead to the resonances in the calculated cross sections that are not observing in the available measurements of the total cross sections and the astrophysical *S*-factor of the proton radiative capture reaction on $^{10}$B at low energies.

Since, the results of the phase shift analysis for the p$^{10}$B elastic scattering are absent and there are not accurate data of the phase shifts, then it might be supposed that the $^6P_{3/2}$ and $^6P_{5/2}$ phase shifts in the energy range less than 1.0 MeV do not need to be equal to zero exactly. They can well be close to zero, i.e., to have the value in order of magnitude of 1–2 degrees. Therefore, let us try to clear one question: can the nonresonance *M*1 transition for certain $^6P_{3/2}$ and $^6P_{5/2}$ potentials of the elastic scattering allow one to correctly reproduce the general shape of the *S*-factor of the proton radiative capture on $^{10}$B above 0.25–0.30 MeV and have the phase shifts close to zero? It was found that the potential parameters the same for both $^6P_{3/2}$ and $^6P_{5/2}$ scattering waves without bound FSs can be represented in the next form

$$V_P = -291 \text{ MeV}, \quad \gamma_P = 1.0 \text{ fm}^{-2}, \qquad (29)$$

which allow, in general, correctly describe the available experimental data on the *S*-factor[133] at the energies from 0.25–0.30 MeV to, approximately, 1.0 MeV, as it was shown below in Fig. 16 by the dashed line. The solid line, as before, shows the summation total cross section for the considered *M*1 and *E*1 transitions (processes No. 1 and No. 7 from Sec. 8.1). The scattering phase shift of this potential at 1.0 MeV reaches 0.8 degree and is shown in Fig. 17. This result quite corresponds to the conception about the closeness of the phase shift to zero and allows one to acceptably reproduce the behavior of total cross sections of the proton capture on $^{10}$B at energies up to 1.0 MeV.

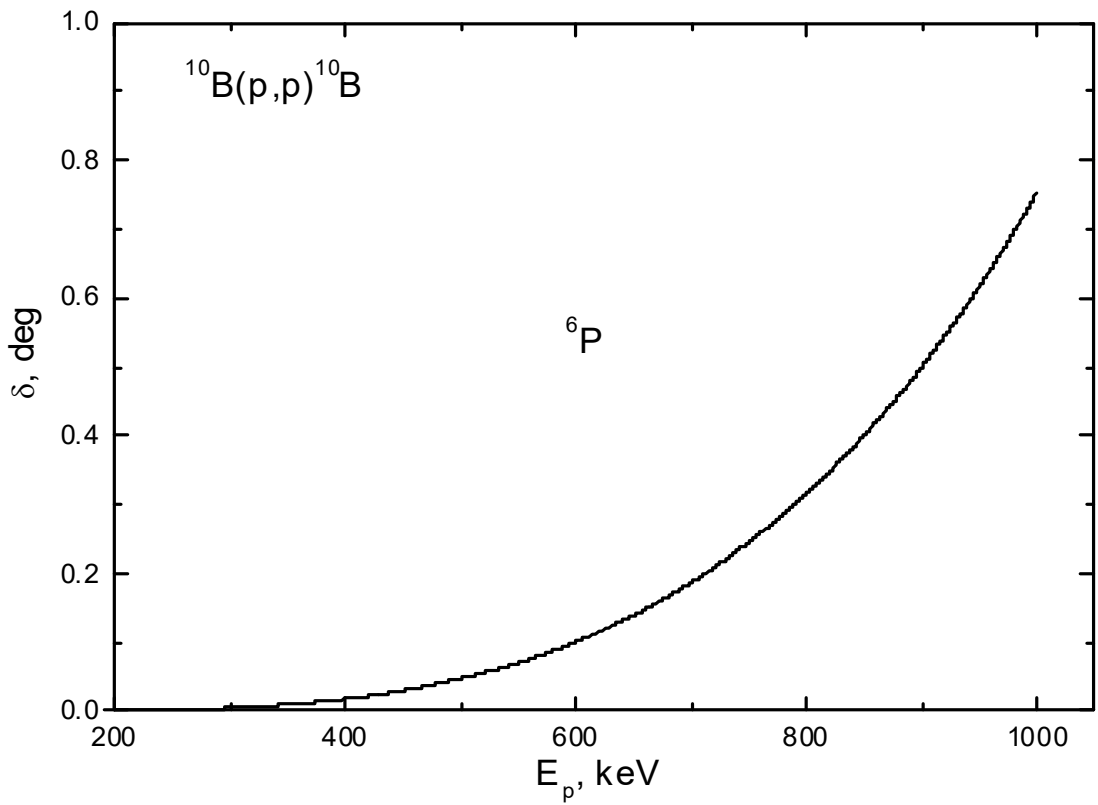

Fig. 17. The $^6P$ phase shift of the p$^{10}$B elastic scattering for the potential (29).

Thereby, it is seen that at the accounting of two $E1$ and $M1$ transitions it is possible, in general, to describe experimental data on the astrophysical $S$-factor or on the total cross sections of the radiative proton capture on $^{10}$B at energies up to 1.0 MeV. And in this system on the basis on more or less simple assumption about the FS and AS structure in the WF of intercluster interaction in the frame of MPCM with FSs it is quite succeeded to reproduce the known experimental measurements.[135]

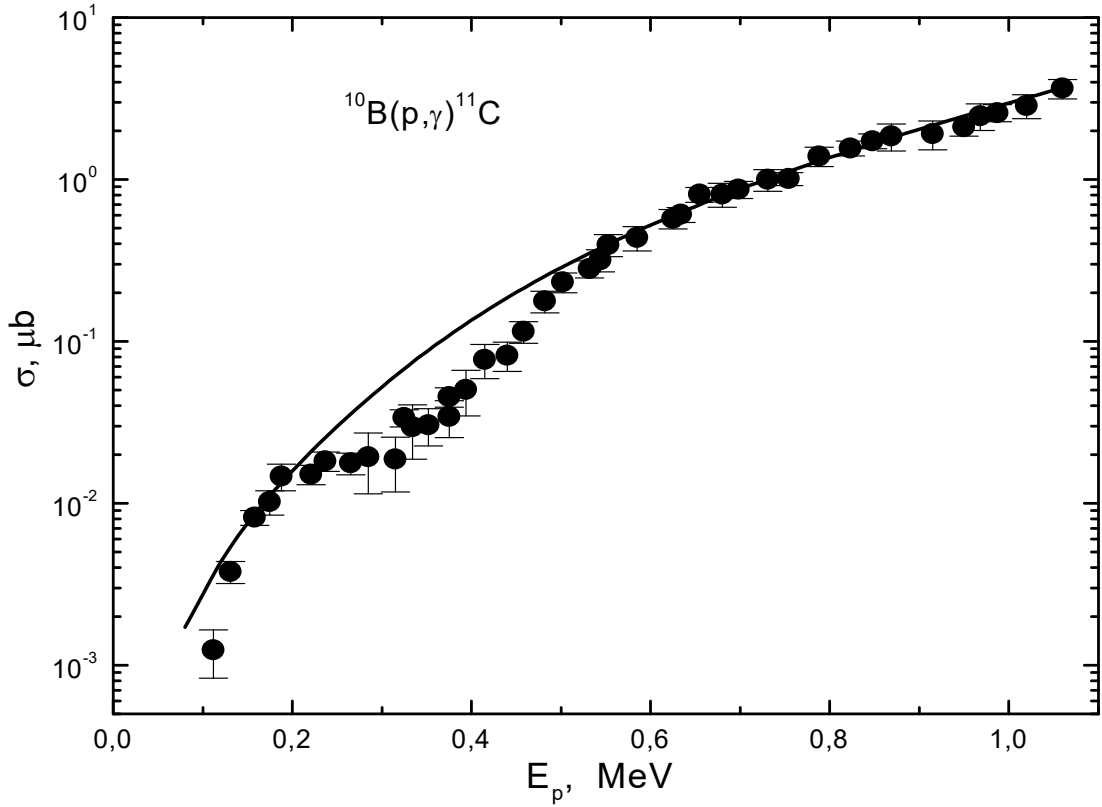

Fig. 18. Total cross sections of the radiative proton capture on $^{10}$B to the GS. Experiment: points (●) – numerical cross section values, obtained in Ref. 133, which were taken from the data base EXFOR, given in the MSU site from Ref. 102.

Furthermore, the total cross section of the radiative proton capture on $^{10}$B is shown in Fig. 18 by the solid line, which corresponds to the astrophysical $S$-factor that shown in Fig. 16. Only the experimental data of Ref. 133 are shown in Fig. 18, and the calculated $S$-factor at lowest energies, as it was seen from Figs. 15 and 16, evidently, better coincide with data of Ref. 134. Generally, the total calculated cross sections are in well agreement with the available experimental data of Ref. 133 at energies more than 400 keV, though the range 200÷400 keV is reproduced slightly worse.

However, the shape of the astrophysical $S$-factor at energies in the range of the first resonance 10 keV did not determined completely at the experimental measurements. The cross section did not measured at the resonance energy; therefore it is prematurely to speak about the correct description of the $S$-factor in the whole range of low energies. At the same cause it is impossible to propose new variant of approximation of the calculated $S$-factor at low energies.

## 7. Proton-capture reaction $^{16}O(p, \gamma)^{17}F$

### 7.1. *Structure of states in the $p^{16}O$ system*

Therefore, for the continuing study of the thermonuclear reactions in the frame of the MPCM with FSs[37,38] let us consider the $^{16}O(p,\gamma)^{17}F$ process, which takes part of the CNO cycle[37] and has additional interest, since it is the reaction at the last nucleus of 1$p$-shell with the forming of $^{17}F$ that get out its limit. As we usually assume,[37,38] the BS of $^{17}F$ is caused by the cluster channel of the initial particles, which take part in the reaction.

Continuing study the total cross sections of the radiative proton capture on $^{16}O$, let us firstly consider classification of orbital states for the $p^{16}O$ system according to Young tableaux. The orbital Young tableau $\{4444\}$[36] corresponds to the ground BS of $^{16}O$, therefore we have $\{1\} \times \{4444\} \to \{5444\} + \{44441\}$[79] for the $p^{16}O$ system. The first of the obtained tableaux compatible with the orbital moment $L = 0$ and is forbidden, because it could not be five nucleons in the $s$-shell, and the second tableau is AS and compatible with the orbital moment $L = 1$.[77] Thereby, in the potential of the $^2S_{1/2}$ wave, which corresponds to the first excited state (FES) of $^{17}F$ at 0.4953 MeV with $J^\pi = 1/2^+$ relative to the GS or -0.1052 MeV relative to the threshold of the $p^{16}O$ channel and scattering states of these particles, there is the forbidden bound state. The $^2P$ scattering waves do not have the bound FSs, and the AS $\{44441\}$ can be located in continuous spectrum. The GS of $^{17}F$ with $J^\pi,T = 5/2^+,1/2$ in the $p^{16}O$ channel, which is the $^2D_{5/2}$ wave at the energy -0.6005 MeV[136] relative to the threshold of the $p^{16}O$ channel and do not have forbidden BSs.

On the basis of data of $^{17}F$ spectrum[136] one can consider that the $E1$ transition is possible from $^2P$ scattering waves with potential without FS for the $^2S_{1/2}$ FES of $^{17}F$ with the bound FS.

$$\text{Process No.1.} \quad \begin{array}{c} ^2P_{1/2} \to {}^2S_{1/2} \\ ^2P_{3/2} \to {}^2S_{1/2} \end{array}.$$

Let us consider the $E1$ transition from the $^2P_{3/2}$ scattering wave with potential without FS for radiative capture to the $^2D_{5/2}$ GS without FS.

$$\text{Process No.2.} \quad {}^2P_{3/2} \to {}^2D_{5/2}.$$

The GS and FES potentials will be constructed so that to describe the channel binding energy and the AC of $^{17}F$ in the $p^{16}O$ channel correctly.

### 7.2. *Phase shift analysis*

Elastic scattering phase shifts of the considered particles usually are used in the frame of the MPCM for the construction of cluster interaction potentials or for nucleons with nuclei. Evidently, one of the first measurements of the differential cross sections of the $p^{16}O$ elastic scattering with carrying out of the phase shift analysis at energies of 2.0–7.6 MeV was done in Ref. 137. This analysis used results of Refs. 138 and 139 and some unpublished results of Ref. 137 in the energy range 2.0–4.26 MeV and 4.25–7.6 MeV, respectively. Furthermore, in Ref. 140, the polarizations of the $p^{16}O$ elastic scattering in the energy range 2.0–5.0 MeV were measured and new phase shift analysis was done, which does not take into account the resonance at 2.66 MeV. Furthermore, in Refs. 141 and 142 the detailed phase shift analysis of the $p^{16}O$ elastic scattering was carried out at energies of 1.5–3.0 MeV and the presence of the narrow resonance was shown for the specified subsequently energy of protons equaling 2.663(7) MeV and the width of 19(1) keV, which corresponds to the first superthreshold state at 3.104 MeV with $J^\pi = 1/2^-$ relative to the GS or 2.5035 (c.m.) above the threshold of the $p^{16}O$ channel.[136] Subsequently, the processes of the elastic scattering were considered in many works (see, for example, Refs. 136,143,144) in the energy range 1.0–3.5 MeV. Particularly, in Refs. 145 and 146 the energies from 0.5–0.6 MeV and to 2.0–2.5 MeV were considered, but the phase shift analysis was not carried for them.

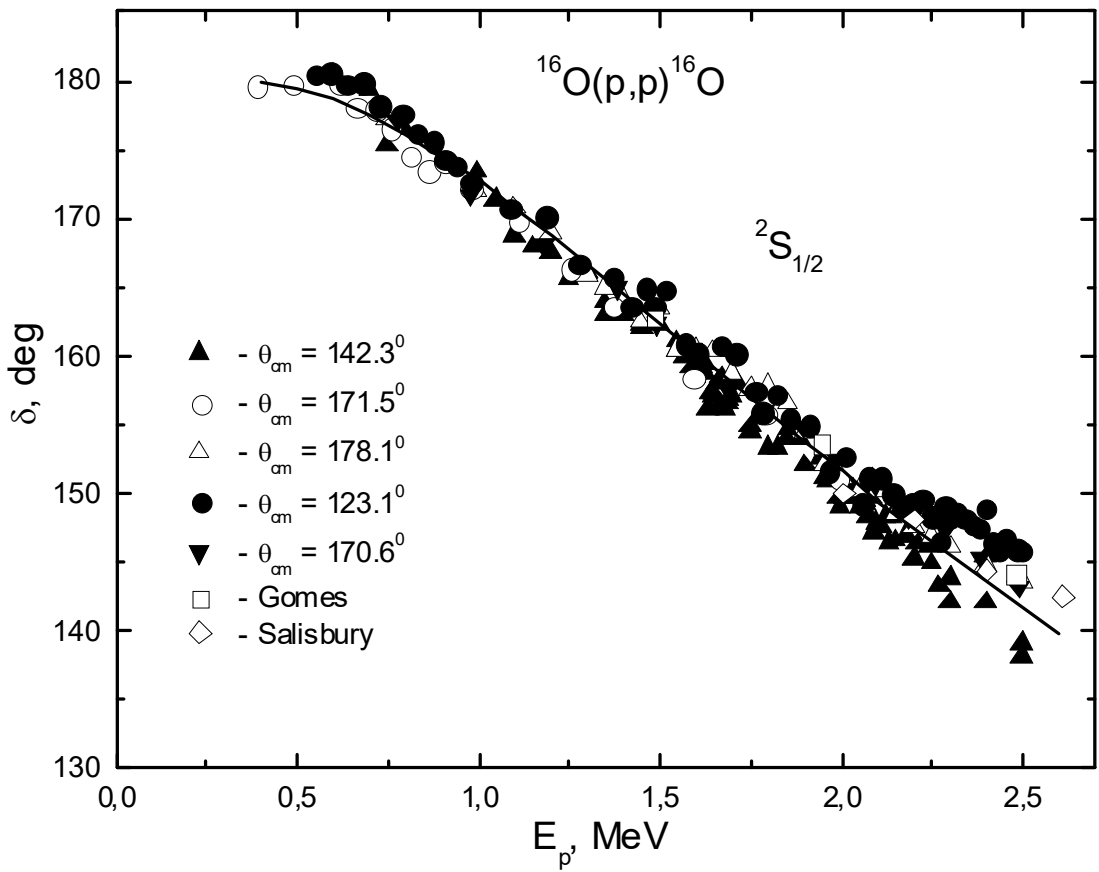

Fig. 19. The $p^{16}O$ scattering phase shifts, obtained in this work on the basis of the excited functions at angles more than 120°. Experimental data are from Ref. 146–149. Open rhombs and squares are results of the analysis at energies from 1.5 MeV to 2.5 MeV.[137,141]

Since, we will further consider the radiative capture in the energy range up to 2.5 MeV, the results of the listed works are quite enough for carrying out of the detailed phase shift analysis at energy, starting from 0.5 MeV. Furthermore, the potentials of the $p^{16}O$ interaction will construct according to this phase shifts up to 2.5 MeV, i.e., without taking into account the first resonance with $J^\pi = 1/2^-$ at 2.663 MeV (l.s.).[136] Thereto, the phase shift analysis of the available data was carried out at the energy range from 0.5 MeV to 2.5 MeV. Meanwhile, it was considered that in this energy range all *P* and *D* scattering phase shifts equal or approximate to zero

Therefore, only one scattering phase shift $^2S_{1/2}$ takes part in our analysis and, since in this wave in the considering energy range there are no resonances, it will have smoothly dropped shape, and results of such analysis are shown in Fig. 19. Experimental data on excitation functions are from Ref. 146–149. Our results[150] are in a good agreement with obtained earlier phase shifts from Refs. 137,141 that are shown in Fig. 19 by squares and rhombs at energies from 1.5 MeV to 2.5 MeV.[137,141]

### 7.3. *Asymptotic constants*

ANC date are given, for example, in Ref. 151. Here we will also use the known relation (4) where *S* is the spectroscopic factor; *C* is the AC in fm$^{-1/2}$, which join with the dimensionless AC $C_W$,[59] using by us further in the next way: $C = \sqrt{2k_0} C_W$, and the dimensionless value $C_W$ defines by the expression (3).[59]

The radius of the GS of $^{16}$O equals 2.710(15) fm from Ref. 136 or 2.6991(52) fm from Ref. 152 was used in further calculations. For the GS and the FES of $^{17}$F the data for radiuses are absent,[136,152] but they, apparently, should not differ a lot from the corresponding data for $^{16}$O. The charged proton radius and its mass radius are equal to 0.8775(51) fm.[87] The accurate values of nucleus and proton masses: $m(^{16}O) = 15.994915$ amu[86] and $m(p) = 1.007276466812$ amu[87] were used in all our calculations.

Table 8. The data on $A_{NC}$ of $^{17}$F in the $p^{16}$O channel and astrophysical S-factors of the proton capture $^{16}$O.

| Value of the $A_{NC}$ in fm$^{-1/2}$ for the GS | Value of the $A_{NC}$ in fm$^{-1/2}$ for the FES | S(0) keV b for the GS | S(0) keV b for the FES | S(0) keV b for the total | References |
|---|---|---|---|---|---|
| 1.59 | 98.2 | --- | --- | 10.37 | 151 |
| 1.04(5) | 75.5(1.5) | 0.40(4) | 9.07(36) | 9.45(40) | 153 |
| 1.04(5) | 80.6(4.2) | 0.40(4) | 9.8(1.0) | 10.2(1.04) | 154 |
| 1.04(5) | --- | 0.317(25) | 8.552(43) | 8.869(44) | 148, 152 |
| 1.10(1) | --- | --- | --- | --- | 148, 155 |
| 1.19(2) | 81.0(9) | --- | --- | 7.1-8.2 | 155 |
| 1.13(1) | 82.3(3) | --- | --- | 7.1-8.2 | 155 |
| 0.97-1.09 | 86.4-91.1 | --- | --- | 10.2-11.0 | 156 |
| 0.97-1.59 | 74.0-98.2 | 0.29-0.44 | 8.7-10.8 | 7.1-11.06 | *Data Limit* |
| 1.28(31) | 86.1(12.1) | 0.37(7) | 9.76(1.04) | 9.08(1.98) | *Average $\bar{A}_{NC}$ per interval* |

Furthermore, we will consider the radiative proton capture on $^{16}$O to the GS of $^{17}$F, which, as it was said, is the $^2D_{5/2}$ level and its potential has to describe the AC correctly. In order to extract this constant from the available experimental data, let us consider information about spectroscopic factors $S$ and $A_{NC}$. The obtained results for $A_{NC}$ are listed in Table 8, in addition, we succeeded in finding a lot of data of spectroscopic factors of $^{17}$F in the $p^{16}$O channel, and therefore we present their values in the form of separate Table 9.

Table 9. Data on the spectroscopic factors $S$ of $^{17}$F in the $p^{16}$O channel of $^{17}$F

| $S$ for the GS | $S$ for the FES | References |
|---|---|---|
| 0.878 | 0.921 | 151 |
| 0.90(15) | 1.00(14) | Results of 157 |
| 0.88 | 0.99 | Is given in 157 with ref. to other works |
| 0.94 | 0.83 | 136 |
| *0.88-1.05* | *0.83-1.14* | *Data Limit* |
| *0.97(9)* | *0.99(15)* | *Average $\bar{S}$ per interval* |

As it is seen from Table 9, the average values of the spectroscopic factors are close to unit, therefore, for easy, we will consider them equals 1. Furthermore, on the basis of expression (4) for the GS we are finding that $\bar{A}_{NC}/\sqrt{\bar{S}} = \bar{C} = 1.28$ fm$^{-1/2}$, and since $\sqrt{2k_0} = 0.57$, so dimensionless AC, determined as $\bar{C}_w = \bar{C}/\sqrt{2k_0}$, is equal to $\bar{C}_w = 2.25$. However, the interval of the $A_{NC}$ values is so much that can lays in the limit 1.7–2.8. For the FES at $\sqrt{2k_0} = 0.37$ we obtain the value $\bar{C}_w = 232.7$ by analogy and the interval of the $C_w$ values taking into account $A_{NC}$ errors equal 200–265. These intervals can be expanded more if we will take into account the errors of the spectroscopic factors $S$ from Table 9.

### 7.4. *Interaction potentials*

For carrying out the calculations of the radiative capture in the frame of the MPCM it is necessary to know potentials of the $p^{16}$O elastic scattering in the $^2P$ waves, and also interactions of the $^2D_{5/2}$ GS and the $^2S_{1/2}$ first excited, but bound in the $p^{16}$O channel, state of $^{17}$F. There are experimental data for total cross sections of the radiative capture exactly for the transition to these BSs that were measured in Refs. 147 and 158. Let us consider further the $E1$ transition to the GS only, but present the potential to the FES of $^{17}$F in the p$^{16}$O channel too, since the phase shift value at energies from 0.5 to 2.5 MeV are known for the $^2S$ wave. For example, for description of the obtained $^2S_{1/2}$ phase shift as a result of our phase shift analysis it is possible to use the simple Gaussian potential with the point-like Coulomb part, FS and parameters of Ref. 38

$$V_S = -87.855325 \text{ MeV}, \quad \gamma_S = 0.15 \text{ fm}^{-2}. \quad (30)$$

Energy dependence of the $^2S_{1/2}$ phase shift of potential (29) is shown in Fig. 19 by the solid line, which starts from 180° because of the presence of the FS.[36] Such potential describes well the behavior of the $^2S$ scattering phase shift, obtained in our analysis, and quite agree with our previous extracting of the phase shifts.[137,141] At the same time, potential (29) allows to obtain the binding energy of the FES equals -0.105200 MeV at the accuracy of the finite-difference method (FDM) used here equals $10^{-6}$ MeV,[130] the charged radius of 3.05 fm, the mass radius of 2.90 fm of $^{17}$F, and the AC of the $p^{16}$O channel at the distance interval of 5–20 fm is equal to $C_W = 197(2)$. This value is in a good agreement with the given above results of its extracting from $A_{NC}$ and spectroscopic factor $S$, laying at the bottom limit of the obtained above interval.

The parameters for the $^2D_{5/2}$ potential of $^{17}$F in the $p^{16}$O channel without FSs were found

$$V_D = -85.632465 \text{ MeV}, \quad \gamma_D = 0.12 \text{ fm}^{-2}, \quad (31)$$

which allow one to obtain the binding energy -0.600500 MeV at the FDM at the accuracy of $10^{-6}$ MeV,[130] the charged radius of 2.82 fm, the mass radius of 2.77 fm, and the AC at the distance interval of 7–28 fm is equal to $C_W = 1.68(1)$ that is also at the lower limit of the given above interval for the AC. The phase shift of this potential drops smoothly from zero and at 2.5 MeV equals the value approximately of -2°.

Since, there are no resonances of the negative parity in spectra of $^{17}$F at the energy lower than 2.5 MeV; we will consider that $^2P$ potentials should to lead in this energy range practically to zero scattering phase shifts, and therefore they does not contain bound BSs, their depth can simple have a zero value.

### 7.5. *Total capture cross sections*

Furthermore, the total cross sections of the proton capture on $^{16}$O were considered on the basis of the MPCM[37,38] and taking into account experimental data.[147,158] Results of our calculations for the $E1$ transitions to the GS with the potential (30) from $^2P$ scattering waves with potentials of zero depth in comparison with experimental data are shown in Fig. 20 by the solid line. The measurement results of the total cross sections in the range from 0.4 MeV to 2.5 MeV[147] are shown by black squares, and circles are data from Ref. 158 at 0.5–2.5 MeV. The error corridor of the experimental data from Ref. 158 is shown by two dashed lines in Fig. 20.

The results of Ref. 153 are shown by the triangles, which lead to the $S$-factor at zero energy of 0.4 keV b. The value of 0.19 keV b at energy of 30 keV was obtained in our calculations, which we can consider the zero energy, because the value of the $S$-factor at energies 30–100 keV changes from 0.19 to 0.21 keV b. The calculated line at energies 0.5–2.5 MeV is in the limit of the available experimental errors of Refs. 147 and 158. However, at lower energy than 0.5–1.0

MeV, it decreases slightly quicker than the existent tendency, which is formed by these experimental data.[158] The linear extrapolation of these data lower 1.0 MeV expressed by formula

$$S(E) = 0.237E(\text{MeV}) + 0.405 \qquad (32)$$

is shown in Fig. 20 by the dotted line and leads to the results at zero energy of 0.405(5) keV b. The dotted line describes data[158] shown in Fig. 20 by circles at energies up to 2.5 MeV quite well, and is in a good agreement with the results[153] at zero energy.

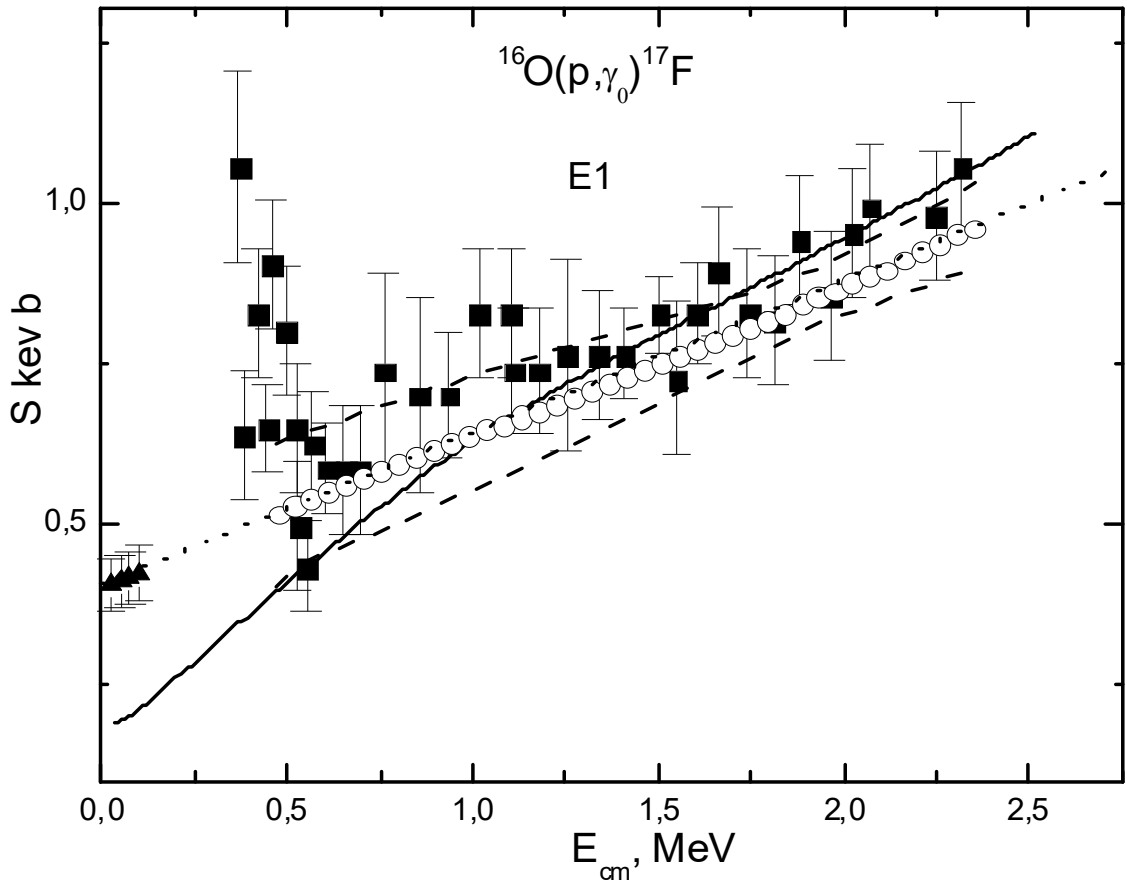

Fig. 20. The total cross section of the proton capture on $^{16}$O to the GS of $^{17}$F. Experimental data: squares – Ref. 147, circles – Ref. 158. Lines are described in the text, triangles – results of Ref. 153.

However, we can see from Fig. 20 that the carried out calculations of the $E1$ transition describe quite well the results of the experimental measurements of the astrophysical $S$-factor from Refs. 147 and 158 to the GS of $^{17}$F, lying in the range of experimental errors. Meanwhile, the potentials of the $^2P$ scattering waves, which do not contain FSs are constructed on the basis of simple assumptions about coordination of such potentials with scattering phase shifts equal, in this case, to zero. The potential of the $^2D_{5/2}$ GS of $^{17}$F in the $p^{16}$O channel was preliminarily coordinated with the basic characteristics of this nucleus, notably, with the binding energy and the AC in the $p^{16}$O channel.

## 8. Conclusions

### 8.1. $^8Li(n,\gamma)^9Li$ reaction

#### 8.1.1 Nuclear astrophysics

Because our calculations[94] do not confirm the large value of the $^8Li(n,\gamma)^9Li$ reaction rate (see Table 4), we can draw a conclusion that the essential part of $^8Li$ will be used for alpha capture with the formation of $^{11}B$. Therefore, the $^8Li(n,\gamma)^9Li$ reaction does not prevent the formation of nuclei with $A > 12$ with the help of the reaction sequence $^8Li(\alpha,n)^{11}B(n,\gamma)^{12}Be(\beta^-)^{12}C(n,\gamma)...$ This means that in the inhomogeneous early Universe, nucleosynthesis in the regions rich in neutrons could produce isotopes with $A > 12$, whereas nucleosynthesis at the standard big bang finishing at the production of $^7Li$.[22] However, the $^8Li(n,\gamma)^9Li$ capture reaction leads nevertheless to a decrease in production elements with $A > 12$. As it was shown by Ref. 4, this process can reduce the production of heavy elements by half. Our results lead to the reaction rate being approximately seven times smaller than Ref. 4, and suggest that the decrease of the formation of heavy elements is not as great.

#### 8.1.2. Nuclear physics

In the frame of the MPCM it is possible to construct the two-body potentials of the n$^8$Li interaction, which allows one to describe the available experimental data for the total cross section of the neutron radiative capture on $^8$Li at low and ultralow energies correctly. Theoretical cross sections are calculated from the thermal energy 25.3 meV to 1.0 MeV and approximated by a simple function of energy, which can be used for calculation of the cross sections at energies up to 50–100 keV. The proposed variants of the GS potentials of $^9$Li in the n$^8$Li channel allow one to obtain the AC in the limits of the available errors and lead to a reasonable description of the $^9$Li radius. Any of the proposed variants of the GS potentials allow one to obtain the calculated capture cross sections that agree reasonably well with available experimental data.[22] The calculation results of the cross sections are not sensitive to the number of the bound FSs and the width of the $^4S$ scattering potential. The obtained results for the total cross sections and static characteristics of $^9$Li depend only on the parameters of the GS potential of this nucleus in the n$^8$Li channel.

### 8.2. $^9Be(p,\gamma)^{10}B$ reaction

Thus, the $E1$ transitions from the $^5S_2$ and $^3S_1$ waves of scattering to the ground $^5P_3$ BS in the p$^9$Be channel of $^{10}$B and its three excited states $1^+0$, $0^+1$ and $1^+0$ also bound in this channel are considered in the MPCM. Having made certain assumptions of common character about interactions in the p$^9$Be channel of $^{10}$B it is possible to describe well the existing experimental data of the astrophysical $S$-factor at the energy range down to 600 keV and to receive its value for zero (50 keV) energy which is in complete agreement with the latest experimental data.[28,104,105] The $S(0)$-factor at zero energy for the transition to the GS of $^{10}$B is obtained generally good.

However, the scattering potentials are constructed on the basis of some qualitative conceptions because of the absence of the data for the phase shift analysis of the p$^9$Be elastic scattering and three BS potentials are obtained only approximately, because of the absence of data on the radii and the AC of $^{10}$B in these excited states. Therefore, these results should be considered only as a preliminary estimate of the possibility to describe the astrophysical *S*-factor of the reaction under consideration on the basis of the MPCM with FSs. But it is possible to obtain quite acceptable results for the astrophysical *S*-factor in spite of approximate consideration of the p$^9$Be → $^{10}$Bγ radiative capture process.

In conclusion, it should be mentioned that the available *S*-factor experimental data for this reaction differs significantly from each other and it seems that more accurate study of the p$^9$Be → $^{10}$Bγ radiative capture at astrophysical energy and more precise definition of the location of the resonance and its width in the $^3S_1$ scattering wave[159] are required.

We should note that in Ref. 31 the variant of classification of orbital states was considered for the n$^9$Be system, when the Young tableau {441} was used for $^9$Be. Consequently, the correct description of the available data of the neutron radiative capture on $^9$Be at astrophysical energies was obtained. Therefore, it is impossible to give the certain preference to one of these variants of state classifications in the n$^9$Be system. However, furthermore we will use for nuclei only the Young tableaux, which concern to their GS, i.e., without taking into account tableaux of FS in these nuclei. For example, the Young tableaux {4411} or {442} will be used for $^{10}$B, since, because of the absence of tables of Young tableau products, it is impossible to unambiguously believe that the last of them is forbidden.

### 8.3. $^{10}$Be(n,γ)$^{11}$Be reaction

In the frame of the MPCM with the classification of states according to Young tableaux, it is quite possible to construct the potentials of the n$^{10}$Be interaction, which allow one to convey the general trend of the experimental data for the total cross sections of the neutron radiative capture on $^{10}$Be at low and ultralow energies generally correctly. The theoretical cross sections are calculated from thermal energy 25.3 meV to 2.0 MeV and are approximated by a simple function of energy, which can be used to calculate the cross sections at energies below 10–50 eV. The proposed variants of the potential of the ground and the first excited states of $^{11}$Be in the n$^{10}$Be channel allow one to obtain the AC in the limit of available for it errors, and lead to a reasonable description of the radius of $^{11}$Be.

### 8.4. $^{10}$B(p,γ)$^{11}$C reaction

Thus, the methods of obtaining the shape and the depth of the intercluster interactions for scattering and BSs that used here allow to get rid of the discrete and continuous ambiguity of its parameters,[130,160] which are inherent to the optical model[161] and observing in the usual approaches at the construction of the intercluster potentials in the continuous and discrete spectrum of the two-body system. Furthermore, potentials obtained by the same way can be used in any calculations connected with the solution of nuclear-physical and astrophysical problems of low, lowest and thermal interaction energies.[36,130,160]

## 8.5. $^{16}O(p,\gamma)^{17}F$ reaction

Thereby, the considered above methods of constructions of interaction potentials of clusters allow one generally correctly reproduce experimental data for total cross sections of the radiative capture at energies from 0.5 MeV to 2.5 MeV. It was obtained the potential of the $^2D_{5/2}$ scattering wave without FS with the phase shift approximate to zero, which is used for the GS and describes its basic characteristics. The potential for the $^2S_{1/2}$ wave with FS correctly reproduces the form of the phase obtained in the result of our phase shift analysis and agrees with the fundamental characteristics of the FES. In conclusion we note that it is thirty cluster systems, which is considered by us on the basis of MPCM with the classification of the orbital states according to states according to Young tableaux, where it is possible to obtain acceptable results on describing characteristics of the radiative capture processes of nucleons or light clusters on nuclei of 1p-shell.

**Acknowledgments**

The work was performed under grant No.0070/GF4 "Thermonuclear reactions in the stars and controlled thermonuclear fusion" of the Ministry of Education and Science of the Republic of Kazakhstan.
   In conclusion, the authors express their deep gratitude to Prof. Rakhim Yarmukhamedov (INP, Tashkent, Uzbekistan) for provision of the information on the ACs in different channels. And Prof. Strakovsky I.I. (GWU, Washington, USA) for discussion of certain issues in the paper.